\pdfoutput=1
\documentclass[conference]{IEEEtran}
\usepackage[T1]{fontenc}
\usepackage[utf8]{inputenc}
\usepackage{tgtermes}
\usepackage{tgheros}
\usepackage[scaled=0.80]{DejaVuSansMono}
\usepackage{amsmath,amssymb}
\usepackage{pifont}
\usepackage{newunicodechar}
\usepackage{graphicx}
\newsavebox{\ucbox}\newdimen\ucwd
\newcommand{\ucell}[1]{\begingroup\ucwd=\fontcharwd\font`0\relax
  \sbox{\ucbox}{\scalebox{0.8}{#1}}%
  \ifdim\wd\ucbox>\ucwd \makebox[\ucwd][c]{\resizebox{\ucwd}{!}{\usebox{\ucbox}}}%
  \else \makebox[\ucwd][c]{\usebox{\ucbox}}\fi\endgroup}
\newcommand{\umath}[1]{\ucell{\ensuremath{#1}}}
\newunicodechar{¬}{\umath{\lnot}}
\newunicodechar{·}{\umath{\cdot}}
\newunicodechar{×}{\umath{\times}}
\newunicodechar{←}{\umath{\leftarrow}}
\newunicodechar{→}{\umath{\rightarrow}}
\newunicodechar{↑}{\umath{\uparrow}}
\newunicodechar{∀}{\umath{\forall}}
\newunicodechar{∃}{\umath{\exists}}
\newunicodechar{∧}{\umath{\land}}
\newunicodechar{∨}{\umath{\lor}}
\newunicodechar{≠}{\umath{\neq}}
\newunicodechar{▸}{\umath{\blacktriangleright}}
\newunicodechar{⟨}{\umath{\langle}}
\newunicodechar{⟩}{\umath{\rangle}}
\newunicodechar{₁}{\umath{{}_1}}
\newunicodechar{₂}{\umath{{}_2}}
\newunicodechar{ℂ}{\umath{\mathbb{C}}}
\newunicodechar{ℕ}{\umath{\mathbb{N}}}
\newunicodechar{Γ}{\umath{\Gamma}}
\newunicodechar{χ}{\umath{\chi}}
\newunicodechar{𝒮}{\umath{\mathcal{S}}}
\newunicodechar{✔}{\ucell{\ding{52}}}
\newunicodechar{✖}{\ucell{\ding{54}}}
\usepackage{booktabs}
\usepackage{array}
\usepackage{xcolor}
\usepackage{fvextra}
\usepackage{tikz}
\usetikzlibrary{positioning,arrows.meta}
\usepackage[hidelinks]{hyperref}
\usepackage{cite}
\usepackage{fancyhdr}
\fvset{fontsize=\fontsize{7.6}{9}\selectfont, breaklines=true, breakanywhere=true,
  breaksymbolleft=\tiny\ensuremath{\hookrightarrow}, breakindent=1em,
  frame=single, framerule=0.3pt, rulecolor=\color{black!40}, framesep=2pt, tabsize=2}
\RecustomVerbatimEnvironment{Verbatim}{Verbatim}{}
\newcommand{\code}[1]{\texttt{#1}}
\newcommand{\RH}{\code{RiemannHypothesis}}
\newcommand{\anyG}{\code{anyGeometry}}

\begin{document}

\title{Does a Trivially True Predicate Invalidate a Kernel-Checked Proof?\\
An Independent Build and Dependency Audit of the Alp\"oge Lean Archive for the Riemann Hypothesis}

\author{\IEEEauthorblockN{Eryk Schiller}
\IEEEauthorblockA{Institute of Industrial and IT Engineering (inTECH)\\
Geneva School of Engineering, Architecture and Landscape (HEPIA)\\
University of Applied Sciences and Arts Western Switzerland (HES-SO)\\
Geneva, Switzerland\\
email: eryk.schiller@hes-so.ch}}

\maketitle
\thispagestyle{fancy}

\begin{abstract}
On 20 November 2026 L.~Alp\"oge released a Lean~4 archive of 175,187 source files whose headline
theorem has type \code{RiemannHypothesis}. A critique by Nielsen argues that the formalization
fails because its geometric witnesses are trivial or degenerate and several associated equalities
hold by definition. In particular, the geometry predicate on the theorem's dependency route is the
constant predicate \code{True}.
We report an independent source build of the archive with its pinned Lean toolchain, against the
prebuilt Mathlib cache for the pinned Mathlib commit. The build completed without error. The
statement is Mathlib's Riemann Hypothesis, and the reported axiom dependencies are exactly
\code{propext}, \code{Classical.choice} and \code{Quot.sound}.
We then examine the compiled environment with two transitive dependency traversals whose resulting
sets agree exactly and contain 1,478,219 constants. We confirm that \anyG{} is reached and that the non-trivial predicate
\code{geoOf} is not. The geometric certificate is constructed and transported, but we detect no
explicit elimination of it: the recursors of the two types that hold it are not reached. The audit
includes detection of primitive projections and was validated on a synthetic development.
These findings separate two questions the critique merges: what the intermediate constructions
mean, and whether the closed formal derivation is valid. Trivially inhabited predicates,
definitional equalities and explicitly assigned intermediate quantities do not, by themselves,
invalidate a theorem accepted by a sound verification chain. We do not establish that the geometric
certificate is removable, we do not identify the mathematical mechanism of the archive, and we have
not traced the other constructions the critique reports. Independent kernel verification has been
reported by others; we have not reproduced those results. Our result addresses the
specific logical objection; it is not a certification of the claimed proof. All commands and their
output are given in the appendix.
\end{abstract}

\begin{IEEEkeywords}
formal verification, Lean 4, Mathlib, proof assistants, Riemann Hypothesis, reproducibility,
dependency analysis
\end{IEEEkeywords}

\section{Introduction}

A proof assistant reduces the question ``is this proof correct?'' to two smaller ones: does the
formal statement say what is claimed, and did a small trusted program, the kernel, accept the
proof term? For a release that claims the Riemann Hypothesis and consists of 175,187 machine-written
files with no accompanying paper, those two questions are the ones an outside reader can answer
soonest. Targeted study of the mathematics is also possible, but it is slower.

Nielsen~\cite{nielsen} answers a third question instead. The critique reads the route from the
top-level theorem downward and finds that a predicate meant to express a geometric condition is
defined as \code{True}. It concludes that the formalization ``certifies only the tautology
`RiemannHypothesis = RiemannHypothesis'\,'' and ``does not contain a proof of the Riemann
Hypothesis''.

This paper reports what we did to test that conclusion. We make three contributions.
\begin{itemize}
\item A full source build of the archive, with the toolchain the release pins and the prebuilt
Mathlib for the pinned revision, and a record of its cost (Section~\ref{sec:build}). We regard this
as a result in its own right, whatever the outcome of the debate.
\item A check of the statement and of the axioms of the compiled theorem
(Section~\ref{sec:statement}).
\item A dependency audit of the compiled proof that follows the \code{True} predicate from the place
where a proof of it is created through the places where it is carried
(Section~\ref{sec:audit}).
\end{itemize}

Our finding is narrow, and it consists of four claims. (i) The formal theorem has the intended type
and the reported axiom dependencies are the standard ones. (ii) The disputed certificate is
constructed and transported, and no elimination of it is detected. (iii) Trivial intermediate
constructions do not imply that a soundly checked derivation is invalid. (iv) Independent kernel
verification has been reported by others~\cite{leanblog,rubinstein}; we have not reproduced those
results or established their precise relationship to the environment audited here. The mathematical
mechanism of the archive is outside the scope of this study.

We do not claim to understand the mathematics of the archive, and we do not claim that the Riemann
Hypothesis is now settled. A kernel can have an implementation defect, imported declarations can be
compromised, and a formal statement can fail to capture the intended claim. Those are real concerns.
The lack of geometric meaning in an internal construction is not, by itself, one of them.

The rebuttal is also independent of how the matter ends. If the archive later turns out to contain a
verification defect, the critique's conclusion would be true for another reason; its argument for
that conclusion would still not be valid.

\section{Background}

\subsection{The archive}

The release is a Lake project named \code{tree}: 175,187 \code{.lean} files, a
\code{lakefile.toml}, a manifest and a toolchain file. It pins Lean v4.33.1~\cite{lean4} and
Mathlib~\cite{mathlib} at commit \code{0df444a3}. The file \code{src/Main.lean} states five theorems.
The first is
\begin{Verbatim}
theorem riemannHypothesis : RiemannHypothesis := rhOfAll
\end{Verbatim}
The other four concern zeros of Dirichlet $L$-functions; their statements are defined in
\code{src/Statements.lean}, which imports only Mathlib.

\subsection{The critique}

Nielsen's critique~\cite{nielsen} makes observations of three kinds.
\begin{enumerate}
\item \emph{The geometry predicate is \code{True}.} The hypothesis bundle on the route to
\code{everyCharacter} is built with \code{geoCl anyGeometry}, where
\code{anyGeometry := fun \_ \_ \_ => True}, and not with the non-trivial predicate \code{geoOf}.
\item \emph{The geometric witness is a degenerate model}, built on \code{geoDegenerate}, with
empty index types and zero Chern numbers.
\item \emph{Lattice and height equalities hold by \code{rfl}}, because one side is defined as a
copy of the other.
\end{enumerate}
From these the critique concludes that the two sides of the final comparison are the same object,
so that the theorem is a tautology. Its Appendix~A anticipates the objection that a kernel-checked
closed term cannot be made false by vacuous intermediate steps, and replies that ``a kernel that
accepts it has checked how the term was assembled, not that the argument holds''.

This paper tests observation~1 at the level of the compiled proof. Observations 2 and 3 were not
examined in the compiled proof; we answer the inference drawn from them in
Section~\ref{sec:height} and list them in Section~\ref{sec:limits}. The critique also names
further structures, among them \code{GeoHalf} and \code{InGeo}, which we did not examine.

\subsection{What a kernel-checked closed term establishes}\label{sec:kernel}

In Lean a proposition is a type and a proof is a term of that type. The theorem
\code{riemannHypothesis} has no hypotheses and no parameters. If a sound kernel accepts a term of
type \RH{}, then \RH{} is derivable in Lean's type theory from the axioms that the term uses. This
holds whatever the internal definitions look like, and it does not require any intermediate
proposition to have mathematical significance of its own. A small example:
\begin{Verbatim}
theorem exampleProof (P : Prop) (h : P) : P := by
  have h₁ : True := trivial
  have h₂ : 2 = 2 := rfl
  exact h
\end{Verbatim}
The two intermediate facts contribute nothing, and the theorem is valid. A second example is
closer to the case at hand, because the theorem is closed and the result passes through a
certificate with a trivial field:
\begin{Verbatim}
structure Certificate (n : Nat) where
  geo : True
  arithmetic : n = 2 + 3
def witness : Certificate 5 := ⟨trivial, rfl⟩
theorem certifiedArithmetic : 5 = 2 + 3 :=
  witness.arithmetic
\end{Verbatim}
The field \code{geo} is trivial. The conclusion comes from the other field, and it is not weakened
by its neighbour.

The critique writes that a kernel ``has checked how the term was assembled, not that the argument
holds''. For a formal proof these are the same thing. The assembly of a well-typed term according
to the typing rules \emph{is} the argument; the rules encode deduction.

The critique's phrase ``a term that is definitionally an identity always type-checks'' also needs
care. A term of type \RH{} is not a proof of \code{RiemannHypothesis = RiemannHypothesis}. That
equation is a different proposition, provable by \code{rfl} for any proposition whatever. The
critique accepts that the top-level term has type Mathlib's \RH{} and concludes that only the
reflexive equation was proved. Under a sound verification chain these two statements cannot both
hold, unless some further defect is identified.

Three separate things are in play, and it helps to keep them apart: derivability in the formal
theory, soundness of the implemented checking process, and faithfulness of the formal statement
to the intended claim. A release of this kind can accordingly fail in several broad ways:
\begin{enumerate}
\item the statement is not the intended one;
\item the kernel was bypassed, for example by an added axiom, \code{sorry}, compiled evaluation
(\code{native\_decide}), or metaprogramming that installs declarations without checking;
\item the kernel, the toolchain, or an imported compiled library has a defect or was altered.
\end{enumerate}
We do not claim the list is exhaustive. The first two classes can be examined directly, and we do
so. The third is addressed by replay. That is outside the scope of this paper: checks of an exported
proof by Lean's kernel and by independent checkers are reported in a joint statement of the Lean FRO
and the Mathlib Initiative~\cite{leanblog} and by Rubinstein~\cite{rubinstein}. We rely on those
reports without having reproduced them (Section~\ref{sec:limits}).

\section{Method}

\subsection{Build}\label{sec:build}

We downloaded the fourteen parts of the archive, joined them, and recorded the SHA-256 digest of the
result. Before extracting it we checked that it contains no absolute paths and no \code{../}
components. The archive contains only source files and three project files; it ships no compiled
\code{.olean} files and no \code{.lake} directory, so every compiled module of the project comes
from our build. The toolchain was installed with \code{elan}, and the prebuilt Mathlib for the pinned
commit was fetched with \code{lake exe cache get}. The Mathlib checkout was verified to be at the
pinned commit and unmodified. Everything else was compiled from source with \code{lake build}.

This is therefore a source build of the archive against cached Mathlib, not a source rebuild of the
whole trusted environment. Mathlib's compiled files were downloaded from the official cache and
are trusted as delivered, as are the Lean binaries installed by \code{elan}.

The build ran on a virtual machine with 20 cores. It was stopped once on purpose, after 19~hours,
to raise the memory from 31~GB to 62~GB, and then resumed; Lake builds are incremental, so no work
was repeated. A side script deleted the per-module \code{setup.json} files once the corresponding
compiled module existed, which keeps disk use near 100~GB instead of about 1~TB.

\subsection{Statement and axioms}\label{sec:statement}

A Lean file that imports \code{Main} was run against the compiled project. It prints the type and
definition of the target, the definition of \RH{}, the module in which \RH{} and
\code{riemannZeta} are declared, and the axioms on which each of the five theorems depends
(\code{\#print axioms}).

\subsection{Source scan for kernel bypasses}

We searched all 175,187 source files for the words \code{axiom}, \code{sorry}, \code{sorryAx},
\code{native\_decide}, \code{unsafe}, \code{opaque}, \code{implemented\_by}, \code{extern},
\code{addDecl}, \code{addDeclWithoutChecking}, \code{skipKernelTC}, \code{ofReduceBool},
\code{reduceBool}, \code{run\_cmd}, \code{elab} and \code{import Lean}. Each file that matched was
read. The scan reports, for each listed word, where it occurs or that it does not occur under the
search pattern used. We inspected the matches. This does not establish that every possible bypass
has been excluded.

\subsection{Dependency audit}\label{sec:audit}

The audit runs inside Lean on the compiled environment. Starting from \code{riemannHypothesis} it
walks through every constant that appears in the type or the value of a reached constant. Proof
terms of theorems are included, so the walk covers the terms the kernel checked, not what the source
text suggests. We call the result the \emph{syntactic transitive dependency closure} of the theorem.
It is the set of constants the proof term mentions, directly or indirectly. It is not the set of
results that are mathematically indispensable. Precisely, the closure is generated by three kinds of
edge: from a constant to every constant in its type; from a definition, theorem or opaque constant
to every constant in its value; and the edges that \code{ConstantInfo.getUsedConstantsAsSet} adds
for constants without a value, namely from an inductive type to its constructors and from a
recursor to the inductive types of its mutual block. A recursor is therefore not
reached merely because its type is reached; it is reached only if some reached term mentions it.

For each reached constant the audit records:
\begin{itemize}
\item whether it refers to \anyG{}, to \code{geoOf}, or to the wrapper \code{geoCl}, and whether
the reference is in its statement (type) or in its proof (value);
\item whether it refers to the field projection \code{OAt.geo}, to the constructor \code{OAt.mk},
or to an eliminator of \code{OAt} (\code{rec}, \code{casesOn}, \code{recOn});
\item whether it refers to a constructor or an eliminator of the inductive wrapper \code{GeoCl}
(\code{base}, \code{step}; \code{rec}, \code{casesOn}, \code{recOn}, \code{below}, \code{brecOn});
\item whether its type or value contains a primitive projection (\code{Expr.proj}) out of
\code{OAt}.
\end{itemize}
It also records a parent for every constant, which gives a shortest chain from the theorem to
\anyG{}.

The reason for watching these names is that they are the ways to use the objects in question. In
Lean, a value of a structure is used by projecting a field or by applying its recursor; a proof of an
inductive proposition is used by applying its recursor, and pattern matching compiles to
\code{casesOn}. If none of these occurs in the closure, no reached proof extracts information from
the object.

\emph{Filtering rules.} In the first audit run (\code{FullAudit.lean}) the user counts leave out
constants that themselves lie in the namespaces \code{OAt} and \code{GeoCl}, because the generated
projections and recursors refer to each other. In that run the projection scan was also applied only
to constants that mention \code{OAt} by name. Both rules could hide a use. The second run
(\code{Audit3.lean}) therefore applies no namespace exclusion and scans every reached constant for
projections without any pre-filter, and it lists every reached constant inside the two namespaces.
The scan tests field indices 0 to 7. \code{OAt} has exactly two fields, \code{geo} (index 0) and
\code{lam} (index 1), and a projection with a larger index is ill-typed and would be rejected by
the kernel, so the bound loses nothing here; it would have to be raised for a structure with more
than eight fields.

\emph{Validation.} The second audit program was first run on a small synthetic development
(\code{Audit3Test.lean}) with the same shape: a structure with a proof field, an inductive wrapper,
and four deliberate uses of them, namely a \code{cases} on the wrapped proof, a pattern match on the
structure, a field read, and a primitive projection inserted directly into a term that does not
mention the structure by name. The program reports all four (Appendix~\ref{app:test}). Two
further cases mark the limits of the method. In a negative control the certificate is built and
carried and only the second field is read; the program reports the constructors and the two field
projections and no eliminator. In the last case a proof of the wrapper is passed to a hypothesis of
the form $P \to Q$. That consumes the proof with no eliminator of the wrapper, and the program, as
expected, reports none. The audit is a targeted syntactic analysis. It is not a complete analysis
of how information flows through a proof.

\emph{Two traversals.} Two walkers were written separately, one depth-first (\code{Check.lean}) and
one breadth-first. Their sets of reached names are compared directly, by a line difference and a
digest of the sorted lists, not only by size. They
share Lean's environment and expression utilities, so they are two traversals of one environment,
not two independent validations of it. Their verdict on \anyG{} and \code{geoOf} was also compared
with that of \code{Name.transitivelyUsedConstants} from the ImportGraph package, which is the check
the critique itself proposes.

\section{Results}

\subsection{Build}

Table~\ref{tab:build} summarizes the build. The log contains no error line, and one compiled module
exists for each of the 175,187 source files.

\begin{table}[t]
\caption{Build of the archive}\label{tab:build}
\centering
\begin{tabular}{@{}ll@{}}
\toprule
Archive SHA-256 & \code{9631b4f0\ldots 3f49c7fc} \\
Source files & 175,187 \\
Lean & 4.33.1, commit \code{819816b2} \\
Mathlib & commit \code{0df444a3}, unmodified \\
Machine & 20 cores; 31~GB, later 62~GB RAM \\
Wall-clock time & 52~h 12~min, in two runs \\
Lake jobs & 183,903, exit code 0 \\
Error lines in log & 0 \\
Compiled modules & 175,187 \\
Build output & 98~GB \\
\bottomrule
\end{tabular}
\end{table}

\subsection{Statement and axioms}

The target and \code{riemannZeta} are both declared in the Mathlib module
\code{NumberTheory.LSeries.RiemannZeta}. The archive does not redefine them. Lean prints the
target as
\begin{Verbatim}[fontsize=\fontsize{6.3}{7.8}\selectfont]
def RiemannHypothesis : Prop :=
∀ (s : ℂ), riemannZeta s = 0 → (¬∃ n, s = -2 * (↑n + 1)) → s ≠ 1 → s.re = 1 / 2
\end{Verbatim}
In words: every zero of the Riemann zeta function that is not one of the trivial zeros
$-2,-4,\dots$ has real part $1/2$, that is, it lies on the line $\tfrac12 + i x$.

All five theorems of \code{Main.lean} depend on exactly \code{propext}, \code{Classical.choice}
and \code{Quot.sound}. These are the standard axioms used throughout Mathlib. The walk reaches
three axioms in total, the same three. As a check that does not go through the name \RH{}, we
also wrote the statement out in full in a new theorem and proved it from \code{riemannHypothesis}:
\begin{Verbatim}[fontsize=\fontsize{6.3}{7.8}\selectfont]
theorem rh_restated :
    ∀ s : ℂ, riemannZeta s = 0 → (¬∃ n : ℕ, s = -2 * ((n : ℂ) + 1)) → s ≠ 1 → s.re = 1 / 2 :=
  riemannHypothesis
\end{Verbatim}
Lean accepts it and reports the same three axioms (Appendix~\ref{app:audit3}). This is strong evidence that the conclusion was not
introduced as an extra axiom. It is a statement about the dependency closure of these theorems. It
is not the stronger statement that every declaration in the environment was soundly constructed;
that rests on trust in the build.

The proof of the target is two steps long at the top:
\begin{Verbatim}[fontsize=\fontsize{6.3}{7.8}\selectfont]
theorem rhOfAll : RiemannHypothesis := Formalism.VW.rh_of_every allChi
theorem allChi : UpperTheory.EveryCharacter := UpperTheory.FinalTheorems.everyCharacter
\end{Verbatim}
where \code{EveryCharacter} states that no Dirichlet $L$-function vanishes for
$\operatorname{Re} s > 1/2$, apart from the trivial character at $s=1$.

\subsection{Source scan}

No file contains \code{axiom}, \code{sorry}, \code{sorryAx}, \code{native\_decide},
\code{unsafe}, \code{opaque}, \code{implemented\_by}, \code{extern}, \code{skipKernelTC},
\code{addDeclWithoutChecking}, \code{ofReduceBool} or \code{reduceBool} as a word. Two calls to
\code{addDecl}, four \code{elab} commands and five \code{run\_cmd} blocks exist
(Appendix~\ref{app:meta} lists every such line). We read each of them. The two \code{addDecl} calls,
in one command \code{cpt\_copy\_thms}, copy existing declarations under new names; \code{addDecl} is
the ordinary entry point that sends a declaration to the kernel. The other three \code{elab} forms
define two tactics and one probe command that elaborate a term and hand it to the goal. The
\code{run\_cmd} lines in \code{BuildAid} add entries to a name deny-list for library suggestions.
We found none that installs a declaration without a kernel check. As said above, this is a finding
about the listed words only.

\subsection{The predicate is on the route}

The critique's factual claim is confirmed. \anyG{} is reached from \code{riemannHypothesis}; the
non-trivial predicate \code{geoOf} is not. The shortest chain has nine constants:
\begin{Verbatim}[fontsize=\fontsize{6.3}{7.8}\selectfont]
riemannHypothesis
rhOfAll
allChi
UpperTheory.FinalTheorems.everyCharacter
UpperTheory.FinalTheorems.everyCharacter_of_row
UpperTheory.LiteratureTheorems.every_character_of_literature_of_vWin
UpperTheory.Lit.every_character_of_theta_integrals_local_matching
GlobalTheory.Mem.Top.Bl.parametersOfDependence
GlobalTheory.Mem.Top.Bl.anyGeometry
\end{Verbatim}
The three theorems the critique names on this route are also reached
(Appendix~\ref{app:full}, lines beginning \code{PROBE}), and so is \code{geoDegenerate}. In total 1,478,219 constants are reached: 1,246,092 theorems, 214,231
definitions, 3 axioms, 91 opaque constants and 17,802 others. The two walkers reach the same set:
the sorted lists of names have the same SHA-256 digest (\code{10f01985\ldots 5fef434e}) and their
line difference is empty. The second walk met no name that is absent from the environment and no
theorem or definition whose body it could not read. Of the 91 opaque constants, 89 are declared in
Mathlib and 2 in Lean's core library; none is declared in the archive
(Appendix~\ref{app:audit3}).

\subsection{No elimination of the geometric certificate is detected}

\begin{figure}[t]
\centering
\begin{tikzpicture}[node distance=5mm, every node/.style={font=\footnotesize},
  box/.style={draw, rounded corners=2pt, align=center, inner sep=3pt, text width=74mm},
  arr/.style={-{Stealth[length=2mm]}}]
\node[box] (a) {\textbf{created}: \code{divEx\_prov}\\ field \code{geo} filled with \code{trivial}, a proof of \code{True}};
\node[box, below=of a] (b) {\textbf{re-wrapped}: \code{divEx\_geoCl}\\ \code{geoCl\_of o.geo} \; (\code{GeoCl.base})};
\node[box, below=of b] (c) {\textbf{carried}: \code{oAt\_setDiv}\\ \code{GeoCl.step N h hp.geo}};
\node[box, below=of c, dashed] (d) {\textbf{no explicit elimination detected}\\ no recursor of \code{OAt} or \code{GeoCl} is reached};
\draw[arr] (a) -- (b); \draw[arr] (b) -- (c); \draw[arr] (c) -- (d);
\end{tikzpicture}
\caption{Life of the proof of the \code{True} predicate on the route to \code{riemannHypothesis}.}
\label{fig:flow}
\end{figure}
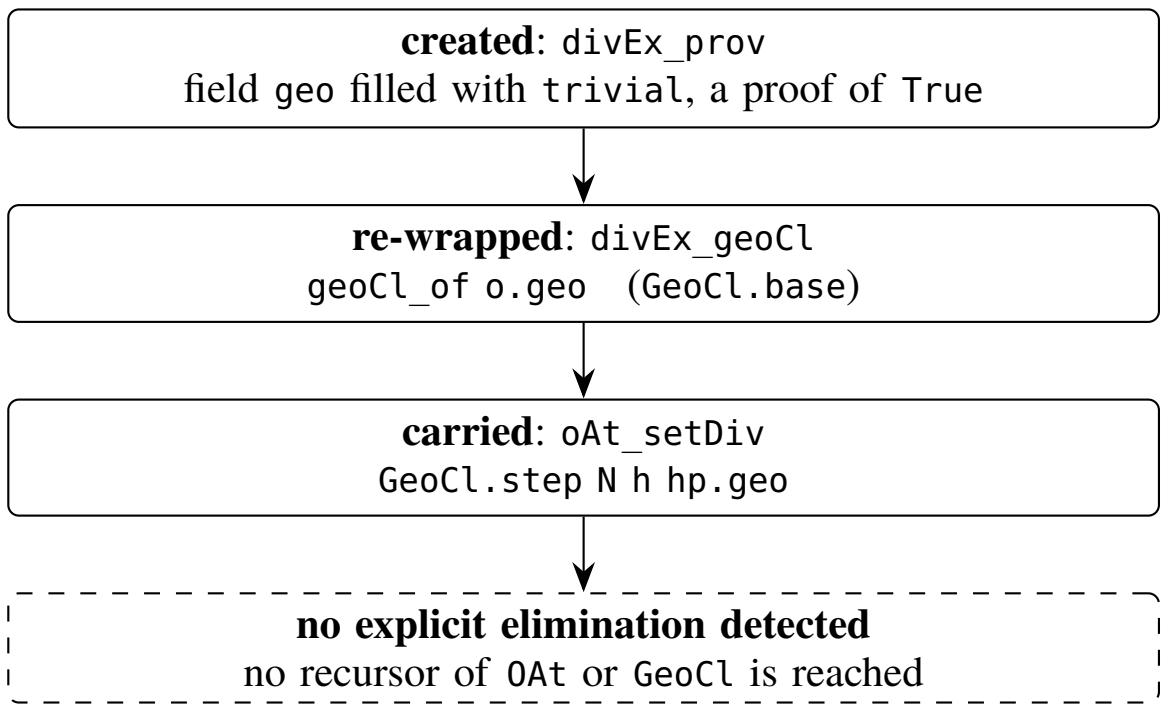

The predicate enters proofs through one structure,
\begin{Verbatim}
structure OAt (Γ : GeoP) ... where
  geo : Γ (w.st Z).toD7 t.r.z.core.toCore X
  lam : LamAre t.r.Z X
\end{Verbatim}
whose field \code{geo} holds a proof that the predicate $\Gamma$ is satisfied, and through one
inductive wrapper, \code{GeoCl} $\Gamma$, with constructors \code{base} and \code{step}.
Table~\ref{tab:trace} gives the number of reached constants that refer to each relevant name, and
Fig.~\ref{fig:flow} shows the resulting flow.

\begin{table}[t]
\caption{Reached constants that refer to each name\\(out of 1,478,219; second run, no namespace excluded; the name itself is not counted)}\label{tab:trace}
\centering
\footnotesize\setlength{\tabcolsep}{3pt}
\begin{tabular}{@{}>{\raggedright\arraybackslash}p{0.40\columnwidth}r>{\raggedright\arraybackslash}p{0.42\columnwidth}@{}}
\toprule
Name & Users & Which \\
\midrule
\code{anyGeometry} & 14 & 10 in the statement, 12 in the proof; Appendix~\ref{app:audit3} \\
\code{geoOf} & 0 & \\
\code{OAt.mk} (build) & 3 & \code{divEx\_prov}, \code{divEx\_geoCl}, \code{oAt\_setDiv} \\
\code{OAt.geo} (read field) & 2 & \code{divEx\_geoCl}, \code{oAt\_setDiv} \\
\code{OAt.lam} (read field) & 3 & the same two and one more (see text) \\
\code{OAt.rec}, \code{casesOn}, \code{recOn} & 0 & \\
primitive projection out of \code{OAt}, any field & 0 & outside \code{OAt.geo} and \code{OAt.lam} themselves \\
\code{GeoCl.base} (build) & 1 & \code{geoCl\_of} \\
\code{GeoCl.step} (build) & 1 & \code{oAt\_setDiv} \\
\code{GeoCl.rec}, \code{casesOn}, \code{recOn} & 0 & \\
\code{GeoCl.below}, \code{brecOn} & 0 & \\
\bottomrule
\end{tabular}
\end{table}

The proof is created once. In \code{divEx\_prov} the field is filled with \code{trivial}:
\begin{Verbatim}
exact ⟨⟨trivial, lamAre_heq hp ht (hX ▸ h4)⟩⟩⟩
\end{Verbatim}
It is read in two places, and in both the value read is immediately placed inside a constructor of
\code{GeoCl} and stored in the \code{geo} field of a new \code{OAt}:
\begin{Verbatim}
-- divEx_geoCl
exact ⟨⟨geoCl_of o.geo, o.lam⟩⟩
-- oAt_setDiv
exact (... (GeoCl.step (Γ := Γ) ... N h hp.geo) :)
\end{Verbatim}
Eliminating a proof of \code{GeoCl} $\Gamma$ directly, by pattern matching or induction, would
introduce a dependency on the elimination principles of \code{GeoCl}. No such dependency is
detected. This does not rule out other uses of the proof: as an argument to a theorem, as a
transported certificate, or as an inhabitance requirement in a dependent construction. The complete list of reached constants in the two
namespaces is
\begin{Verbatim}
GlobalTheory.Mem.Top.Bl.GeoCl
GlobalTheory.Mem.Top.Bl.GeoCl.base
GlobalTheory.Mem.Top.Bl.GeoCl.step
GlobalTheory.Mem.Top.Bl.OAt
GlobalTheory.Mem.Top.Bl.OAt.geo
GlobalTheory.Mem.Top.Bl.OAt.lam
GlobalTheory.Mem.Top.Bl.OAt.mk
\end{Verbatim}
Seven names: the two types, their constructors and the two field projections. No eliminator.
The second run, which excludes no namespace, lists the same seven names. Its projection scan, applied
to all 1,478,219 constants without a pre-filter, finds a primitive projection out of \code{OAt} in
exactly two constants, the generated projection functions \code{OAt.geo} and \code{OAt.lam}
themselves. So no reached term reads a field of \code{OAt} except by calling those two functions,
and the callers of \code{OAt.geo} are the two constants shown above. All uses of \code{OAt.geo},
\code{OAt.mk}, \code{GeoCl.base} and \code{GeoCl.step} are in proofs, none in a statement.

So the \code{True} proof is produced, wrapped and moved, and we detect no point at which
information is explicitly extracted from it. The second field, \code{lam}, states the
lattice-related predicate \code{LamAre}. Unlike \code{geo}, it is read by a third constant,
\code{parameter\_of\_discrete\_series\_ray\_alt}, which does not read \code{geo}. Our audit does
not determine the mathematical strength of this lattice predicate or of its proof. It shows only
that the structure is read downstream for this field and not for \code{geo}:
a trivially inhabited \code{geo} field does not make \code{lam}, or the structure as a whole,
trivial.

It is worth being exact about what this shows. Three levels should be kept apart.
\begin{enumerate}
\item \emph{Syntactic reachability:} established. \anyG{} and the certificate are in the closure.
\item \emph{Explicit consumption:} not detected. No recursor and no primitive projection extracts
the certificate.
\item \emph{Removability:} not established. That the proof would still go through with the field
removed is a claim about a modified development, and we have not built one.
\end{enumerate}
Dependent types complicate ordinary data-flow reasoning. A proof field may be needed merely to
construct an inhabitant of a structure, even if nothing later examines it; strengthening the field
could make the structure impossible to build. So the absence of elimination is not a complete
proof of semantic independence. The logical rebuttal in Section~\ref{sec:kernel} does not depend on
it.

One point can be made without the audit. With $\Gamma$ instantiated to \code{geoCl anyGeometry},
the field \code{geo} is a proposition that has a proof for every argument. Nothing can be deduced
from such a proposition beyond what is already known, so a proof that carries it along gains no
information from it.

\section{Discussion}

Table~\ref{tab:claims} sets the critique's statements beside what we observed.

\begin{table}[t]
\caption{The critique's statements and our observations}\label{tab:claims}
\centering
\begin{tabular}{@{}>{\raggedright\arraybackslash}p{0.43\columnwidth}>{\raggedright\arraybackslash}p{0.47\columnwidth}@{}}
\toprule
Statement in~\cite{nielsen} & Observation \\
\midrule
The theorem inhabits Mathlib's \RH{}; no \code{axiom}, \code{sorry}, \code{native\_decide} & Confirmed (type and axiom report; textual scan of the source) \\
\anyG{} is \code{True} and is on the route & Confirmed \\
\code{geoOf} is not on the route & Confirmed \\
\code{geoDegenerate} is on the route & Confirmed (reached); its use was not traced \\
Lattice and height equalities hold by \code{rfl} & Not examined \\
Downstream ``receives a certificate that is true by construction'' & Confirmed; no explicit elimination of that certificate is detected \\
The kernel certified only ``\RH{} = \RH{}'' & Not so: the checked term has type \RH{} \\
The release contains no proof of the Riemann Hypothesis & Does not follow from the above \\
\bottomrule
\end{tabular}
\end{table}

The critique is a careful reading of the source, and the principal source-level observations that
we examined are confirmed. Its error is in the step from ``this component is trivial'' to ``the
theorem is void''.

A conditional theorem can be vacuously true if its antecedent is unsatisfiable:
$\mathrm{False} \to P$ is provable for every $P$. That possibility does not apply here. The final
theorem has type \RH{} and carries no geometric premise. Moreover, replacing an antecedent by
\code{True} does not weaken an implication: $\mathrm{True} \to P$ is logically equivalent to $P$.
A premise that is trivially satisfied is easy to supply, and what is then obtained is the full
conclusion. So trivially inhabited intermediate conditions cannot, by themselves, trivialize a closed
theorem whose statement does not depend on them.

The critique's Appendix~A says that the kernel objection ``assumes that removing the vacuous inputs
leaves an independent derivation behind''. It does not need that assumption. Soundness of the type
theory says that an accepted closed term of type $P$ establishes $P$; it says nothing about how
the term is organised, and it does not require any part of the term to be meaningful on its own.
The audit adds an empirical observation: for the predicate, no elimination of the certificate is
detected. Whether the derivation survives removal of the field is a separate question
(Section~\ref{sec:removal}).

\subsection{Definitional equalities and intermediate quantities}\label{sec:height}

The critique's remaining objections concern the construction of the divisor datum, the copying of
lattice fields, and the assignment of the height function to an explicit formula. We have not traced
these constructions through the compiled dependency graph. The inference from their definitional
character to the invalidity of the theorem is, however, no better justified than for the predicate.

An intermediate quantity used in an inequality need not have an origin independent of the
quantities compared. If $A(x) \le H(x)$ and $H(x) \le B(x)$ are established, then $A(x) \le B(x)$,
whether $H$ comes from an independent geometric construction or is defined directly in terms of
$x$. Comparison functions, majorants, potentials and Lyapunov functions are routinely chosen to make
an inequality provable.

A small example shows that the principle ``the intermediate quantity must be independent'' is
false. Define $H(x,y) := (x^2+y^2)/2$. Since $(x-y)^2 \ge 0$ we have $xy \le H(x,y)$, and by
definition $H(x,y) \le (x^2+y^2)/2$. Hence $2xy \le x^2+y^2$. Here the intermediate quantity was
chosen to fit, one of the two inequalities is reflexive, and all quantities depend on the same
inputs. The proof is valid.

The critique displays the two inequalities of the archive in the form
\[
\lVert \text{pairing} \rVert \le h, \qquad \mu + h - F_o \le \mathrm{MAIN} + \text{error},
\]
from which $\mu + \lVert \text{pairing} \rVert - F_o \le \mathrm{MAIN} + \text{error}$ follows.
This step is valid however $h$ is defined, provided both inequalities hold under the relevant
premises. Defining $h$ by an explicit formula makes its defining equation immediate. It does not
make the inequalities that involve $h$ immediate, it does not imply
$\lVert \text{pairing} \rVert = h$, and it does not make the comparison reflexive. Showing any of
those would need a separate demonstration, which the critique does not give. The question that
matters is whether the two inequalities are established from permissible premises, and that is
what the kernel checks, within the supplied formal environment. Whether those inequalities are
faithful to the intended geometry and analysis is a separate question.

The point can be put as a small proposition. For real numbers $a,b,c,d$, there exists $h$ with
$a \le h$ and $c + h \le b + d$ if and only if $c + a \le b + d$: one direction is transitivity,
the other takes $h = a$. So the existence of an intermediate quantity supplies no independent
mathematical object, and none is needed. What matters is how the inequalities are established.

So the absence of an independent geometric interpretation of the height would be a limitation of
the geometric account that the archive's names suggest. It would not show that the resulting
inequalities, or the theorem derived from them, are invalid. To show that, one would have to exhibit
an incorrect formal statement, an unjustified dependency, or a failure of the checking process.
This subsection is an argument about logic. Our computational findings remain restricted to the
geometry predicate.

\subsection{What the critique does establish}

What the critique does show is something different and worth saying plainly. It identifies
constructions whose formal content is weaker than the geometric interpretation their names suggest.
For the predicate we confirm this. We do not claim that geometry is absent everywhere in the
archive; we examined one predicate and its wrapper. A reader who
wants to learn \emph{why} the Riemann Hypothesis holds will not learn it from a predicate equal
to \code{True}. That is a statement about how well the archive explains itself. It is not a
statement about whether the theorem is proved.

There is a real tension here, and we do not resolve it. If the geometric conditions on the route
are as weak as the critique reports and the verification chain is sound, then the mathematical content that forces the zeros onto the line must sit
somewhere else in the 1.48~million constants. We have not located it. Until somebody does, the
status of the release is ``accepted by the Lean 4.33.1 kernel, not understood''.

\section{Limitations}\label{sec:limits}

\begin{enumerate}
\item \emph{No replay of our own.} The proof was checked by the kernel of the same Lean binary
that elaborated it, during the build. Three further checks exist and they are not interchangeable.
Replaying the compiled declarations through the pinned kernel
(\code{lean4checker}~\cite{lean4checker}) addresses whether declarations were installed without
ordinary checking. Checking the proof with a separately implemented kernel protects against defects
specific to one implementation; kernels have had soundness bugs before~\cite{lean4lean}. Tools that
export or compare proofs, such as Comparator~\cite{comparator}, add the question of whether the
export is faithful. We did none of the three, because others report them. The joint statement of
the Lean FRO and the Mathlib Initiative~\cite{leanblog} reports that the library compiles, that the
theorem was compared with an isolated statement importing only Mathlib by two tools, one of them
Comparator, and that an exported proof was checked by Lean's kernel and by several independent
checkers. Rubinstein~\cite{rubinstein} reports a build without errors, the same three axioms for
the five theorems, a kernel replay that accepted all 175,187 modules, and a run of the independent
kernel \code{nanoda} over 1,484,729 declarations without a type-checking error. We have not
reproduced these results and cite them as reported.

Two qualifications apply. First, we have not established that those checks ran on the same artifact
as ours. An equal number of files does not show it; that would need a comparison of the archive
digest, the Lean and Mathlib revisions, the target and the checker versions. Our digests are in
Appendix~\ref{app:meta}. Second, the 1,484,729 declarations of the reported \code{nanoda} run and
the 1,478,219 constants of our closure differ by 6,510. The tools may count different targets or
different categories of declaration. We have not established the relation between the two numbers;
they neither contradict nor cross-validate each other.

The critique mentions a Comparator run under a different Lean version and discusses a rejection. We
have not examined it. A rejection could come from incompatibility, an exporter problem, resource
limits, unsupported declarations, or a genuine verification failure, and neither manuscript
establishes which. The statement in~\cite{leanblog} reports a successful Comparator comparison and
does not mention a rejection. We do not dismiss the critique's report.
\item \emph{Cached dependencies.} Mathlib's compiled files came from the official cache and the Lean
binaries from the official installer. Both are part of the trusted base of this report.
\item \emph{One of three constructions traced.} We followed the predicate. We did not follow the
degenerate model (\code{geoDegenerate}) or the equalities proved by \code{rfl}. The arguments of
Sections~\ref{sec:kernel} and~\ref{sec:height} apply to them, but the instrumented evidence in this
paper does not.
\item \emph{Reachability is not necessity.} The audit establishes syntactic reachability and the
absence of detected elimination. It does not establish that the predicate is semantically
independent of the result, or removable.
\item \emph{Users of \anyG{} read in source only.} That the fourteen constants mentioning
\anyG{} merely pass it along as a parameter is shown by the audit for the certificate
(\code{OAt.geo}, \code{GeoCl}); for the name \anyG{} itself it rests on reading the source. The
second run shows where the name occurs (ten statements, twelve proofs) but not what each proof does
with it. Since \anyG{} unfolds to \code{True}, unfolding it yields no information either.
\item \emph{Mathlib is trusted.} We checked that the target is Mathlib's definition, not that
Mathlib's definition of the zeta function is free of error. It is widely used and reviewed.
\item \emph{The source scan is textual.} It finds the listed words; it would not find a bypass
that avoids all of them.
\item \emph{No mathematical account.} We have not located the estimates that carry the argument.
\item \emph{One machine, one build.} The digests of the archive, the manifest and the Lean binary
are given so that others can check that they build the same input.
\end{enumerate}

\subsection{An experiment on removability}\label{sec:removal}

Changing the field to \code{geo : True} in the definition of \code{OAt}, and adjusting the three
proofs that build or move it (\code{divEx\_geoCl}, \code{oAt\_setDiv}, and the unused
\code{oAt\_geoOf}), removes the predicate from every certificate. If the archive still compiles to a
proof of \RH{} after that edit, the stronger interface of the original field is shown not to be
needed for the modified derivation. A failed rebuild would show less: changes to dependent types can
break proofs for unrelated reasons. We have not run this experiment; it requires recompiling every
module downstream of the definition. A study of the estimates that feed
\code{zeroFreeOn\_of\_tiedWin\_cyc} is more informative and should come first.

\section{Conclusion}

We built the Alp\"oge archive from source with its pinned Lean toolchain and cached Mathlib. The
build completed without error. The compiled theorem has the type of Mathlib's \RH{}: every
non-trivial zero of the zeta function has real part $1/2$. Its reported axiom dependencies are the
three standard axioms of classical Lean mathematics.

The dependency audit confirms the critique's principal observation about the geometry predicate.
\anyG{} is the constant predicate \code{True}, it occurs in the transitive dependency closure of
the theorem, and it is used in place of the non-trivial predicate \code{geoOf}. The audit also finds
no explicit elimination of the geometric certificate through the recursors of \code{OAt} or
\code{GeoCl}. The certificate is constructed and transported, and no information is detected as
being extracted from it. This is supported by an unfiltered projection scan and by validation of the
audit program on a synthetic development. It does not establish semantic independence or
removability.

These observations separate two questions that the critique merges. The first is whether the
archive provides the geometric interpretation that its definitions and theorem names suggest. The
second is whether it contains a valid formal derivation of the stated theorem. A negative answer to
the first does not imply a negative answer to the second. In the same way, definitional equalities
and explicitly assigned intermediate quantities do not, by themselves, make later inequalities
reflexive or the final theorem a tautology.

\enlargethispage{-0.59\textheight}
The criticism has substance: it identifies constructions whose formal content is weaker than the
geometric interpretation suggested by their names. That is a legitimate observation about the
archive. It does not establish the invalidity of the theorem. This paper rebuts
an argument for invalidity; it does not exclude every source of invalidity. Our results do not
establish the soundness of the whole verification chain independently of the original Lean
implementation, and they do not identify the mathematics responsible for the zero-free estimates.
Independent verification has been reported by others~\cite{leanblog,rubinstein}; we have not
reproduced it. A mathematical account of those estimates remains open.

The contribution is deliberately restricted: the argument for invalidity that rests on trivial
geometric predicates and definitional constructions is insufficient. Whether the archive
establishes the Riemann Hypothesis has to be judged by the soundness and faithfulness of its
complete formal derivation, not by the mathematical significance of its intermediate names.

The defensible position is therefore not that the Riemann Hypothesis has been proved. It is that
the critique has not shown that it has not.

\clearpage
\onecolumn
\appendices
\fvset{fontsize=\fontsize{8.6}{10.2}\selectfont}

\section{Commands and Their Output}\label{app:cmd}

Each listing shows the commands as run, each on a line that starts with \code{\$}, followed by the
output. Listings are copied from the files in \code{paper/artifacts/}. Where lines are left out this
is said. Long lines are wrapped; a wrapped continuation is marked with $\hookrightarrow$. Host
commands were run on the build machine in the project directory
\code{\textasciitilde/riemann/extracted/lean}; times are UTC.

\subsection{Obtaining and checking the archive}

The parts were fetched and joined with
\begin{Verbatim}
$ for i in $(seq -w 1 14); do curl -fL --retry 5 -sS -O "https://alpo.ge/lean.tar.gz.part$i"; done
$ cat lean.tar.gz.part{01..14} > lean.tar.gz
$ gzip -t lean.tar.gz && echo GZIP_OK
$ sha256sum lean.tar.gz
$ if tar -tzf lean.tar.gz | grep -qE '(^/|\.\./)'; then echo UNSAFE_PATHS; exit 1; fi
$ tar -xzf lean.tar.gz -C ~/riemann/extracted
\end{Verbatim}
in this order: the integrity and path checks ran before extraction. The same checks, repeated
afterwards with their output (file \code{a1-archive.txt}):
\begin{Verbatim}
$ cd ~/riemann/download && sha256sum lean.tar.gz
9631b4f020199e580ba86178cb56c6e9b09f690c2369a80dc4334aa93f49c7fc  lean.tar.gz
$ cd ~/riemann/download && gzip -t lean.tar.gz && echo GZIP_OK
GZIP_OK
$ cd ~/riemann/download && tar -tzf lean.tar.gz | grep -cE '(^/|\.\./)'
0
$ find ~/riemann/extracted/lean/src -name '*.lean' | wc -l
175187
$ cat ~/riemann/extracted/lean/lean-toolchain
leanprover/lean4:v4.33.1
$ grep -A3 'name = "mathlib"' ~/riemann/extracted/lean/lakefile.toml
name = "mathlib"
git = "https://github.com/leanprover-community/mathlib4"
rev = "0df444a360eaa60ab8c11dca51a86af692955474"

\end{Verbatim}

\subsection{Toolchain, Mathlib and build}

The toolchain and the Mathlib cache were installed with
\begin{Verbatim}
$ curl -sSfL https://raw.githubusercontent.com/leanprover/elan/master/elan-init.sh -o elan-init.sh
$ sh elan-init.sh -y --default-toolchain none
$ cd ~/riemann/extracted/lean && lake exe cache get
$ tmux new-session -d -s rh "bash ~/riemann/build.sh"
\end{Verbatim}
The build script (\code{build.sh}) is:
\begin{Verbatim}
#!/bin/bash
# Builds the Lean project. Incremental: re-running resumes where it left off.
# - stops the build if free disk drops below 6 GB
# - deletes per-module *.setup.json (~6 MB each) once that module's .olean exists
export PATH="$HOME/.elan/bin:$PATH"
PROJ=$HOME/riemann/extracted/lean
B=$PROJ/.lake/build
LOG=$HOME/riemann/build.log
MIN_FREE_KB=$((6 * 1024 * 1024))

cd "$PROJ" || exit 2
echo "=== build started $(date -Is) ===" >> "$LOG"
lake build >> "$LOG" 2>&1 &
BUILD=$!
n=0
while kill -0 "$BUILD" 2>/dev/null; do
  free_kb=$(df --output=avail -k "$PROJ" | tail -1)
  if [ "$free_kb" -lt "$MIN_FREE_KB" ]; then
    echo "=== DISK GUARD: only ${free_kb} KB free, stopping build $(date -Is) ===" >> "$LOG"
    kill -TERM "$BUILD"; sleep 5
    for p in $(pgrep -u "$USER" -x lean) $(pgrep -u "$USER" -x lake); do kill -KILL "$p" 2>/dev/null; done
    exit 3
  fi
  n=$((n+1))
  if [ $((n % 3)) -eq 0 ] && [ -d "$B/ir" ]; then
    find "$B/ir" -name '*.setup.json' -mmin +3 -print0 2>/dev/null | while IFS= read -r -d '' f; do
      rel=${f#$B/ir/}; rel=${rel%.setup.json}
      [ -f "$B/lib/lean/$rel.olean" ] && rm -f "$f"
    done
  fi
  sleep 20
done
wait "$BUILD"; rc=$?
echo "=== build finished rc=$rc $(date -Is) ===" >> "$LOG"
exit $rc
\end{Verbatim}
State of the host after the build (file \code{a2-build.txt}). The line \code{rc=143} is the
deliberate stop to add memory; the build was then resumed and ended with \code{rc=0}.
\begin{Verbatim}
$ sha256sum ~/riemann/download/lean.tar.gz
9631b4f020199e580ba86178cb56c6e9b09f690c2369a80dc4334aa93f49c7fc  /home/eryk/riemann/download/lean.tar.gz
$ cd ~/riemann/extracted/lean && lake env lean --version
Lean (version 4.33.1, x86_64-unknown-linux-gnu, commit 819816b2e0a3bf405af45ae5c7af2491d8f5bee6, Release)
$ cd ~/riemann/extracted/lean && git -C .lake/packages/mathlib rev-parse HEAD
0df444a360eaa60ab8c11dca51a86af692955474
$ cd ~/riemann/extracted/lean && git -C .lake/packages/mathlib status --short | wc -l
0
$ grep '^===' ~/riemann/build.log
=== build started 2026-11-21T01:17:19+00:00 ===
=== MANUAL STOP requested (to add RAM) 2026-11-21T20:23:57+00:00 ===
=== build finished rc=143 2026-11-21T20:24:05+00:00 ===
=== build started 2026-11-21T20:30:58+00:00 ===
=== build finished rc=0 2026-11-23T05:36:11+00:00 ===
$ grep -cE '^(✖|error)' ~/riemann/build.log
0
$ tail -n 7 ~/riemann/build.log | cut -c1-100
✔ [183896/183903] Built Development.AutomorphicArithmetic.AutomorphicVarieties.SmallClassNumbers.C
✔ [183898/183903] Built Supplements.Kernels.MainTheorems (590s)
✔ [183899/183903] Built Supplements.Kernels.MainTheoremsCritical (64s)
✔ [183900/183903] Built Supplements.Kernels.MainTheoremsAll (54s)
✔ [183902/183903] Built Main (75s)
Build completed successfully (183903 jobs).
=== build finished rc=0 2026-11-23T05:36:11+00:00 ===
$ find ~/riemann/extracted/lean/.lake/build/lib -name '*.olean' | wc -l
175187
$ nproc; free -g | sed -n 2p
20
Mem:              62           1           1           0          60          61
\end{Verbatim}

\subsection{Statement, axioms and first dependency walk}

Program \code{Check.lean}:
\begin{Verbatim}
import Main

open Lean

#print axioms riemannHypothesis
#print axioms riemannZeta_eq_zero_re
#print axioms LFunction_ne_zero_of_half_lt_re
#print axioms LFunction_eq_zero_re
#print axioms LFunction_eq_zero_re_of_isPrimitive

#check (riemannHypothesis : RiemannHypothesis)
#print RiemannHypothesis

/-- All constants reachable from `root` through types and values. -/
def usedConsts (root : Name) : CoreM NameSet := do
  let env ← getEnv
  let mut seen : NameSet := {}
  let mut stack : Array Name := #[root]
  while !stack.isEmpty do
    let n := stack.back!
    stack := stack.pop
    if seen.contains n then continue
    seen := seen.insert n
    if let some ci := env.find? n then
      let mut cs := ci.type.getUsedConstants
      if let some v := ci.value? (allowOpaque := true) then cs := cs ++ v.getUsedConstants
      match ci with
      | .inductInfo i => cs := cs ++ i.ctors.toArray
      | _ => pure ()
      for c in cs do
        unless seen.contains c do stack := stack.push c
  return seen

def watch : List String :=
  ["anyGeometry", "geoOf", "geoCl", "GeoCl", "plugAll", "shapeWin", "zeroFreeAll_of_topG",
   "zero_free_all_of_green_v_win", "nonempty_hyps_definite_side_of_closed_bound",
   "every_character_of_green_v_win", "every_character_of_theta_integrals_local_matching",
   "every_character_of_literature_of_vWin", "everyCharacter_of_row", "everyCharacter",
   "galoisLocalGlobalStrong_recParameter", "recTExists", "vwin_of_row", "geoDegenerate",
   "hExpl6", "setDiv", "divEx_prov", "exists_divT", "sorryAx", "geoTied", "formsTieX",
   "shape_closed_link_n_of_green", "linkDIVt_closedK", "topDesWin", "zeroFreeOn_of_tiedWin_cyc",
   "rh_of_every", "rh_of_right", "casselmanWallach", "clozelPurity", "bhattScholzeIntegral"]

#eval show CoreM Unit from do
  let env ← getEnv
  let used ← usedConsts ``riemannHypothesis
  let names := env.header.moduleNames
  let mut mods : NameSet := {}
  let mut perLib : Std.HashMap String Nat := {}
  let mut constPerLib : Std.HashMap String Nat := {}
  let mut axioms : Array Name := #[]
  let mut hits : Array String := #[]
  let mut thms := 0
  let mut defs := 0
  for n in used do
    if let some ci := env.find? n then
      match ci with
      | .axiomInfo _ => axioms := axioms.push n
      | .thmInfo _ => thms := thms + 1
      | .defnInfo _ => defs := defs + 1
      | _ => pure ()
    let last := match n with | .str _ s => s | _ => ""
    if watch.contains last then
      let m := match env.getModuleIdxFor? n with | some i => toString names[i.toNat]! | none => "?"
      hits := hits.push s!"{n}  [{m}]"
    if let some i := env.getModuleIdxFor? n then
      let m := names[i.toNat]!
      let lib := toString m.getRoot
      constPerLib := constPerLib.insert lib (constPerLib.getD lib 0 + 1)
      unless mods.contains m do
        mods := mods.insert m
        perLib := perLib.insert lib (perLib.getD lib 0 + 1)
  IO.println s!"TOTAL constants used: {used.size}  (theorems {thms}, definitions {defs})"
  IO.println s!"AXIOM constants reached: {axioms}"
  IO.println s!"MODULES used: {mods.size} of {names.size} imported"
  IO.println "modules used per library:"
  for (k, v) in perLib.toArray.qsort (fun a b => a.2 > b.2) do IO.println s!"  {k}: {v}"
  IO.println "constants used per library:"
  for (k, v) in constPerLib.toArray.qsort (fun a b => a.2 > b.2) do IO.println s!"  {k}: {v}"
  IO.println "WATCHED names on the route:"
  for h in hits.qsort (· < ·) do IO.println s!"  {h}"
  let modList := (mods.toArray.map toString).qsort (· < ·)
  IO.FS.writeFile "/home/eryk/riemann/checks/used_modules.txt" (String.intercalate "\n" modList.toList ++ "\n")
  let constList := (used.toArray.map toString).qsort (· < ·)
  IO.FS.writeFile "/home/eryk/riemann/checks/used_constants.txt" (String.intercalate "\n" constList.toList ++ "\n")
\end{Verbatim}
Command:
\begin{Verbatim}
$ cd ~/riemann/extracted/lean
$ ( date -u -Is; /usr/bin/time -v lake env lean --tstack=400000 -DmaxHeartbeats=0 ~/riemann/checks/Check.lean; echo EXIT=$?; date -u -Is ) > ~/riemann/checks/check.out 2>&1
\end{Verbatim}
Output (\code{check.out}, complete):
\begin{Verbatim}
2026-11-23T08:13:07+00:00
'riemannHypothesis' depends on axioms: [propext, Classical.choice, Quot.sound]
'riemannZeta_eq_zero_re' depends on axioms: [propext, Classical.choice, Quot.sound]
'LFunction_ne_zero_of_half_lt_re' depends on axioms: [propext, Classical.choice, Quot.sound]
'LFunction_eq_zero_re' depends on axioms: [propext, Classical.choice, Quot.sound]
'LFunction_eq_zero_re_of_isPrimitive' depends on axioms: [propext, Classical.choice, Quot.sound]
riemannHypothesis : RiemannHypothesis
def RiemannHypothesis : Prop :=
∀ (s : ℂ), riemannZeta s = 0 → (¬∃ n, s = -2 * (↑n + 1)) → s ≠ 1 → s.re = 1 / 2
TOTAL constants used: 1478219  (theorems 1246092, definitions 214231)
AXIOM constants reached: #[Quot.sound, Classical.choice, propext]
MODULES used: 151691 of 186200 imported
modules used per library:
  Development: 119090
  Formalism: 14927
  GroundTheory: 6578
  Mathlib: 5081
  ReciprocityTheory: 3458
  CompanionTheory: 1882
  PositivityTheory: 343
  Lemmata: 157
  Init: 152
  Batteries: 15
  Std: 4
  Aesop: 1
  Lean: 1
  Supplements: 1
  Main: 1
constants used per library:
  Development: 842309
  Formalism: 278273
  Mathlib: 203035
  GroundTheory: 76752
  ReciprocityTheory: 47612
  CompanionTheory: 15717
  Init: 9662
  PositivityTheory: 4276
  Lemmata: 314
  Batteries: 253
  Std: 9
  Aesop: 3
  Supplements: 2
  Lean: 1
  Main: 1
WATCHED names on the route:
  Formalism.BasicNotions.DV.hExpl6  [Formalism.LocalRingCardinality.KernelPlaceMeanBound]
  Formalism.HC.zeroFreeOn_of_tiedWin_cyc  [Formalism.SharpRestriction.RamificationFrobenius.TotientGaloisClasses.TopWin]
  Formalism.L2.Geo.geoDegenerate  [Formalism.LocalToGlobal.LatticeGeometry.Inhabited]
  Formalism.VW.rh_of_every  [Formalism.GammaFactorNonvanishing.CritLine]
  Formalism.VW.rh_of_right  [Formalism.GammaFactorNonvanishing.RhOfRight]
  GlobalTheory.BSP.bhattScholzeIntegral  [Development.AbelianVarietyTheory.Proofs.InertiaEigenvalues.BhattScholzeIntegral]
  GlobalTheory.EX.OB.setDiv  [Development.AbelianVarietyTheory.Definitions.GlobalTwisting.SetUnipotentTranslate]
  GlobalTheory.Fc.S0.UnitaryGroupThirdPlaceHypotheses.casselmanWallach  [Development.AutomorphicArithmetic.ThetaCorrespondence.ThetaPairLifts.ThirdPlaceSeries.BernsteinUnitaryProductIrreducible]
  GlobalTheory.FunPr.ThetaLocalInputs.casselmanWallach  [Development.AutomorphicArithmetic.ThetaCorrespondence.ThetaPairLifts.LocalTypeDictionary.BernsteinUnitaryProductIrreducible]
  GlobalTheory.Mem.Top.Bl.GeoCl  [Development.AbelianVarietyTheory.Definitions.AutomorphicMembership.NotGeoClFalse]
  GlobalTheory.Mem.Top.Bl.anyGeometry  [Development.AbelianVarietyTheory.Definitions.AutomorphicMembership.OverIx]
  GlobalTheory.Mem.Top.Bl.geoCl  [Development.AbelianVarietyTheory.Definitions.AutomorphicMembership.NotGeoClFalse]
  GlobalTheory.Mem.Top.DivRec.divEx_prov  [Development.AbelianVarietyTheory.Proofs.IwasawaDecomposition.TupPinsOfMuIsW]
  GlobalTheory.Mem.Top.DivRec.exists_divT  [Development.AbelianVarietyTheory.Proofs.IwasawaDecomposition.TupPinsOfMuIsW]
  PositivityTheory.CWB.casselmanWallach  [Development.WeilBound.CoreStatements.PublishedTheorems.CasselmanWallach.FinalB]
  PositivityTheory.Lit1.ClozelPurity.clozelPurity  [Development.WeilBound.CoreStatements.PublishedTheorems.ClozelPurity.Bridge]
  UpperTheory.FinalTheorems.everyCharacter  [Development.AutomorphicArithmetic.CharacterBoxes.MainConclusion]
  UpperTheory.FinalTheorems.everyCharacter_of_row  [Development.AutomorphicArithmetic.CharacterBoxes.MainConclusion]
  UpperTheory.Gal.Int.galoisLocalGlobalStrong_recParameter  [Development.AutomorphicArithmetic.Foundations.GaloisRepresentations.IntegralStructures.RowSrc]
  UpperTheory.Lit.every_character_of_theta_integrals_local_matching  [Development.AutomorphicArithmetic.ThetaCorrespondence.ThetaPairLifts.ThirdPlaceSeries.SiegelEisensteinSecondTermHypotheses]
  UpperTheory.Lit.nonempty_hyps_definite_side_of_closed_bound  [Development.AutomorphicArithmetic.ThetaCorrespondence.ThetaPairLifts.ThirdPlaceSeries.SiegelEisensteinSecondTermHypotheses]
  UpperTheory.LiteratureTheorems.every_character_of_literature_of_vWin  [Development.AutomorphicArithmetic.ThetaCorrespondence.ThetaPairLifts.ThirdPlaceSeries.EveryCharacterOfLiteratureOfVWin]
  UpperTheory.One.vwin_of_row  [Development.AutomorphicArithmetic.AutomorphicVarieties.ModularityArgument.ThresholdArguments.DenseThresholdCases.VwinOfRow]
  UpperTheory.Top.every_character_of_green_v_win  [Development.AutomorphicArithmetic.ThetaCorrespondence.ThetaPairLifts.ThirdPlaceSeries.TopPlug]
  UpperTheory.Top.shape_closed_link_n_of_green  [Development.AutomorphicArithmetic.ThetaCorrespondence.ThetaPairLifts.ThirdPlaceSeries.TopShape]
  UpperTheory.Top.topDesWin  [Development.AutomorphicArithmetic.CharacterBoxes.DesWin]
  UpperTheory.Top.zeroFreeAll_of_topG  [Development.AutomorphicArithmetic.CharacterBoxes.ZeroFreeAllOfTopG]
  UpperTheory.Top.zero_free_all_of_green_v_win  [Development.AutomorphicArithmetic.ThetaCorrespondence.ThetaPairLifts.ThirdPlaceSeries.TopPlug]
  UpperTheory.Zero.LLC.recTExists  [Development.AutomorphicArithmetic.Foundations.LocalLanglands.LocalCentralUnits.RecTExists]
	Command being timed: "lake env lean --tstack=400000 -DmaxHeartbeats=0 /home/eryk/riemann/checks/Check.lean"
	User time (seconds): 2455.60
	System time (seconds): 344.28
	Percent of CPU this job got: 31%
	Elapsed (wall clock) time (h:mm:ss or m:ss): 2:28:35
	Average shared text size (kbytes): 0
	Average unshared data size (kbytes): 0
	Average stack size (kbytes): 0
	Average total size (kbytes): 0
	Maximum resident set size (kbytes): 60342728
	Average resident set size (kbytes): 0
	Major (requiring I/O) page faults: 974674
	Minor (reclaiming a frame) page faults: 4353560
	Voluntary context switches: 1111219
	Involuntary context switches: 6576
	Swaps: 0
	File system inputs: 184347136
	File system outputs: 178312
	Socket messages sent: 0
	Socket messages received: 0
	Signals delivered: 0
	Page size (bytes): 4096
	Exit status: 0
EXIT=0
2026-11-23T10:41:42+00:00
\end{Verbatim}

\subsection{The exact statement and the top of the proof}\label{app:target}

Program \code{TargetAudit.lean}:
\begin{Verbatim}
import Main
open Lean

/-!
Full audit of the Lean Riemann Hypothesis theorem.

Checks:
1. The final theorem and its target
2. The exact RiemannHypothesis definition
3. The theorem's axiom dependencies
4. The geometry definitions
5. The allChi proof route
6. The transitive dependency graph
-/

-- ==========================================
-- 1. FINAL THEOREM
-- ==========================================

#check riemannHypothesis
#check (riemannHypothesis : RiemannHypothesis)

#print riemannHypothesis
#print rhOfAll

-- ==========================================
-- 2. EXACT RIEMANN HYPOTHESIS DEFINITION
-- ==========================================

#print RiemannHypothesis

-- ==========================================
-- 3. AXIOM AUDIT
-- ==========================================

#print axioms riemannHypothesis
#print axioms rhOfAll
#print axioms allChi

-- ==========================================
-- 4. GEOMETRIC DEFINITIONS
-- ==========================================

#print GlobalTheory.Mem.Top.Bl.anyGeometry
#print GlobalTheory.Mem.Top.Bl.geoOf

-- ==========================================
-- 5. FINAL PROOF ROUTE
-- ==========================================

#check allChi
#print allChi

-- ==========================================
-- 6. TRANSITIVE DEPENDENCY AUDIT
-- ==========================================

#eval show CoreM Unit from do
  let used ← (``riemannHypothesis).transitivelyUsedConstants

  IO.println "===== DEPENDENCY RESULTS ====="

  IO.println s!"anyGeometry: {used.contains ``GlobalTheory.Mem.Top.Bl.anyGeometry}"
  IO.println s!"geoOf:       {used.contains ``GlobalTheory.Mem.Top.Bl.geoOf}"

  IO.println "===== END OF AUDIT ====="
\end{Verbatim}
Command:
\begin{Verbatim}
$ cd ~/riemann/extracted/lean
$ lake env lean --tstack=400000 -DmaxHeartbeats=0 /home/eryk/riemann/checks/TargetAudit.lean 2>&1 | tee /home/eryk/riemann/checks/target-audit.out
\end{Verbatim}
Output (\code{target-audit.out}, complete). The
line beginning \code{def RiemannHypothesis} is the statement that all non-trivial zeros have real part
$1/2$; the line \code{anyGeometry: true} is the critique's own proposed check
(\code{transitivelyUsedConstants}) and confirms its factual claim.
\begin{Verbatim}
riemannHypothesis : RiemannHypothesis
riemannHypothesis : RiemannHypothesis
theorem riemannHypothesis : RiemannHypothesis :=
rhOfAll
theorem rhOfAll : RiemannHypothesis :=
Formalism.VW.rh_of_every allChi
def RiemannHypothesis : Prop :=
∀ (s : ℂ), riemannZeta s = 0 → (¬∃ n, s = -2 * (↑n + 1)) → s ≠ 1 → s.re = 1 / 2
'riemannHypothesis' depends on axioms: [propext, Classical.choice, Quot.sound]
'rhOfAll' depends on axioms: [propext, Classical.choice, Quot.sound]
'allChi' depends on axioms: [propext, Classical.choice, Quot.sound]
def GlobalTheory.Mem.Top.Bl.anyGeometry : GlobalTheory.Mem.Top.Bl.GeoP :=
fun x x_1 x_2 => True
def GlobalTheory.Mem.Top.Bl.geoOf : ((D : Formalism.BasicNotions.D7) → GlobalTheory.ShFam.KitClass D) →
  GlobalTheory.Mem.Top.Bl.GeoP :=
fun Q D J X => (∃ c, 0 < c ∧ GlobalTheory.ShFam.IsGeoOf (Q D) J c X) ∧ GlobalTheory.ShFam.Pad.HasTrueSurf X
allChi : UpperTheory.EveryCharacter
theorem allChi : UpperTheory.EveryCharacter :=
UpperTheory.FinalTheorems.everyCharacter
===== DEPENDENCY RESULTS =====
anyGeometry: true
geoOf:       false
===== END OF AUDIT =====
\end{Verbatim}

\subsection{Source scan for kernel bypasses}

File \code{a3-scan.txt}. Words that do not appear in the count do not occur in any source file.
\begin{Verbatim}
$ cd ~/riemann/extracted/lean/src && grep -rhoE --include=*.lean '\b(axiom|sorry|sorryAx|native_decide|unsafe|opaque|implemented_by|extern|addDecl|addDeclWithoutChecking|skipKernelTC|ofReduceBool|reduceBool|run_cmd|elab|import Lean)\b' . | sort | uniq -c | sort -rn
     12 import Lean
      5 run_cmd
      4 elab
      2 addDecl
$ cd ~/riemann/extracted/lean/src && grep -rlE --include=*.lean '\b(addDecl|run_cmd)\b' . | sort
./BuildAid/D1.lean
./BuildAid/D2.lean
./BuildAid/D3.lean
./BuildAid/D4.lean
./BuildAid/D5.lean
./Formalism/HermitianCurvature/HermitianCurvatureProofs/CutCptCopy.lean
\end{Verbatim}

\subsection{Metaprogramming in the source}\label{app:meta}

File \code{a6-meta.txt}: every source line containing \code{elab}, \code{addDecl} or
\code{run\_cmd} as a word.
\begin{Verbatim}
$ cd ~/riemann/extracted/lean/src
$ grep -rnE --include=*.lean '\b(elab|addDecl|run_cmd)\b' . | cut -c1-230
./Formalism/LocalToGlobal/IndefiniteReal/HeckeNonvanishing/Imports.lean:5:elab "kexactB " e:term : tactic => withMainContext do
./Formalism/LocalToGlobal/NumberFieldFamilies/OwnStruct.lean:53:elab "kcopy " t:term : tactic => withMainContext do
./Formalism/LocalToGlobal/NumberFieldFamilies/OwnStruct.lean:60:elab "#kcopy_probe " t:term " : " ty:term : command => Command.runTermElabM fun _ => do
./Formalism/HermitianCurvature/HermitianCurvatureProofs/CutCptCopy.lean:78:      addDecl (.defnDecl { d with name := n', type := subst tainted d.type, value := subst tainted d.value })
./Formalism/HermitianCurvature/HermitianCurvatureProofs/CutCptCopy.lean:82:      addDecl (.thmDecl { t with name := n', type := subst tainted t.type, value := subst tainted t.value })
./Formalism/HermitianCurvature/HermitianCurvatureProofs/CutCptCopy.lean:87:elab "cpt_copy_thms" : command => do
./BuildAid/D3.lean:4:run_cmd Lean.modifyEnv fun env => Lean.LibrarySuggestions.nameDenyListExt.addEntry env "GlobalTheory"
./BuildAid/D2.lean:4:run_cmd Lean.modifyEnv fun env => Lean.LibrarySuggestions.nameDenyListExt.addEntry env "CrystallineSemistableComparison"
./BuildAid/D1.lean:4:run_cmd Lean.modifyEnv fun env => Lean.LibrarySuggestions.nameDenyListExt.addEntry env "CompanionTheory"
./BuildAid/D4.lean:4:run_cmd Lean.modifyEnv fun env => Lean.LibrarySuggestions.nameDenyListExt.addEntry env "PositivityTheory"
./BuildAid/D5.lean:4:run_cmd Lean.modifyEnv fun env => Lean.LibrarySuggestions.nameDenyListExt.addEntry env "UpperTheory"
\end{Verbatim}

\subsection{Digests of the project files and the Lean binaries}

File \code{a7-hashes.txt}. The last two commands show that the archive ships no compiled files: it
contains the sources under \code{lean/src/} and three project files.
\begin{Verbatim}
$ cd ~/riemann/extracted/lean && sha256sum lake-manifest.json lakefile.toml lean-toolchain
fee57ee9ed399d0a638ca955e5ad40a37a36eeb9b6dfd9f5964114fa7bbe4774  lake-manifest.json
b831ad8cac0babb7bf713433259bdfe5163e58c90734e86d71536d4e30ddec4c  lakefile.toml
3aac669c7a910ec2389f4e4f921b605adf6ebf2d1e0c9b9cd0be4d33f3f5db71  lean-toolchain
$ cd ~/riemann/extracted/lean && sha256sum $(lake env which lean) $(lake env which leanchecker)
e8baaa71855a616dc351028f3ad2200051b0671f423a1696a100e809302d5550  /home/eryk/.elan/toolchains/leanprover--lean4---v4.33.1/bin/lean
84bc3e7cd6efa5bfecd1e7c37571fed9428fb2b3e16da0ddf62f2163573486d0  /home/eryk/.elan/toolchains/leanprover--lean4---v4.33.1/bin/leanchecker
$ tar -tzf ~/riemann/download/lean.tar.gz | grep -cE '\.olean|/\.lake/'
0
$ tar -tzf ~/riemann/download/lean.tar.gz | grep -vE '^lean/src/'
lean/lake-manifest.json
lean/lakefile.toml
lean/lean-toolchain
\end{Verbatim}

\subsection{The predicate and the certificate in the source}

File \code{a4-source.txt}. Line 19 of \code{OverIx.lean} is the definition the critique quotes.
The last command shows that \code{oAt\_any} and \code{oAt\_geoOf} occur only where they are
declared.
\begin{Verbatim}
$ cd ~/riemann/extracted/lean/src
$ grep -rn --include=*.lean anyGeometry . | wc -l
118
$ grep -n -E '^(abbrev GeoP|def geoOf|def anyGeometry|  geo :|  lam :)' Development/AbelianVarietyTheory/Definitions/AutomorphicMembership/OverIx.lean
14:abbrev GeoP : Type 1 := ∀ D : D7, Formalism.L2.Pj.Core D → DivX D → Prop
16:def geoOf (Q : ∀ D : D7, GlobalTheory.ShFam.KitClass D) : GeoP := fun D J X =>
19:def anyGeometry : GeoP := fun _ _ _ => True
46:  geo : Γ (w.st Z).toD7 t.r.z.core.toCore X
47:  lam : LamAre t.r.Z X
$ grep -n -E '^(inductive GeoCl|  \| base|def geoCl|theorem geoCl_of)' Development/AbelianVarietyTheory/Definitions/AutomorphicMembership/NotGeoClFalse.lean
21:inductive GeoCl (Γ : GeoP) : ∀ D : D7, Formalism.L2.Pj.Core D → DivX D → Prop
22:  | base {D : D7} {J : Formalism.L2.Pj.Core D} {X : DivX D} : Γ D J X → GeoCl Γ D J X
26:def geoCl (Γ : GeoP) : GeoP := GeoCl Γ
28:theorem geoCl_of {Γ : GeoP} {D : D7} {J : Formalism.L2.Pj.Core D} {X : DivX D} (h : Γ D J X) : geoCl Γ D J X := GeoCl.base h
$ sed -n '204,209p' Development/AbelianVarietyTheory/Proofs/IwasawaDecomposition/TupPinsOfMuIsW.lean
theorem divEx_prov (𝒮 : Slots) (hOx : 𝒮.Ox = OX anyGeometry) : DivEx 𝒮 :=
  StepW.of_all fun _ _ _ _ w _ _ hpin Z l =>
    let ⟨Y, h1, h2, h3, h4⟩ := exists_divT w hpin.some Z l
    ⟨Y, h1, h2, h3, fun O t hX hp ht => by
      rw [hOx]
      exact ⟨⟨trivial, lamAre_heq hp ht (hX ▸ h4)⟩⟩⟩
$ sed -n '49p' Development/AbelianVarietyTheory/Proofs/IwasawaDecomposition/PublishedInputsImplyConclusion.lean
  exact ⟨⟨geoCl_of o.geo, o.lam⟩⟩
$ sed -n '106p' Development/AbelianVarietyTheory/Proofs/IwasawaDecomposition/DivisionGeometryScalarsTransferSetDivision.lean
  · exact (by with_unfolding_all exact ((GeoCl.step (Γ := Γ) (D := (w.st Z).toD7) N h hp.geo) :))
$ grep -rn --include=*.lean -E '\b(oAt_any|oAt_geoOf)\b' . | cut -d: -f1,2
./Development/AbelianVarietyTheory/Definitions/AutomorphicMembership/NotLamAreShift.lean:152
./Development/AbelianVarietyTheory/Definitions/AutomorphicMembership/NotLamAreShift.lean:157
\end{Verbatim}

\subsection{Full dependency audit of the compiled proof}\label{app:full}

The program \code{FullAudit.lean} is given in Appendix~\ref{app:prog}. Command:
\begin{Verbatim}
$ cd ~/riemann/extracted/lean
$ tmux new-session -d -s full "export PATH=\$HOME/.elan/bin:\$PATH; date -u > ~/riemann/checks/full-audit.out; lake env lean ~/riemann/checks/FullAudit.lean >> ~/riemann/checks/full-audit.out 2>&1; echo \"=== exit \$? \$(date -u)\" >> ~/riemann/checks/full-audit.out"
\end{Verbatim}
Standard output (\code{full-audit.out}, complete):
\begin{Verbatim}
Tue Nov 24 12:34:31 PM UTC 2026
Formalism.VW.rh_of_every : UpperTheory.EveryCharacter → RiemannHypothesis
def UpperTheory.EveryCharacter : Prop :=
∀ (N : ℕ) [inst : NeZero N] (χ : DirichletCharacter ℂ N) (s : ℂ),
  1 / 2 < s.re → χ ≠ 1 ∨ s ≠ 1 → DirichletCharacter.LFunction χ s ≠ 0
structure GlobalTheory.Mem.Top.Bl.OAt {K : GlobalTheory.Mem.Top.Ctx} {B : GlobalTheory.Mem.Base}
  {f : GlobalTheory.Mem.Top.Fr B.toPre} {T : GlobalTheory.Mem.Top.TwDat B} (Γ : GlobalTheory.Mem.Top.Bl.GeoP)
  {w : GlobalTheory.Mem.Top.Bl.ColW B f T} {Z : w.Lt} {Wt : Set B.places.Pl}
  {p0 : Formalism.BasicNotions.BaseLevelRowSide (w.D Z)} (t : GlobalTheory.Mem.Top.Bl.RowsT K w Z Wt p0)
  (X : Formalism.BasicNotions.DivX (w.D Z)) : Type 1
number of parameters: 11
fields:
  GlobalTheory.Mem.Top.Bl.OAt.geo : Γ (w.st Z).toD7 t.r.z.core.toCore X
  GlobalTheory.Mem.Top.Bl.OAt.lam : GlobalTheory.Mem.Top.Bl.LamAre t.r.Z X
constructor:
  GlobalTheory.Mem.Top.Bl.OAt.mk {K : GlobalTheory.Mem.Top.Ctx} {B : GlobalTheory.Mem.Base}
    {f : GlobalTheory.Mem.Top.Fr B.toPre} {T : GlobalTheory.Mem.Top.TwDat B} {Γ : GlobalTheory.Mem.Top.Bl.GeoP}
    {w : GlobalTheory.Mem.Top.Bl.ColW B f T} {Z : w.Lt} {Wt : Set B.places.Pl}
    {p0 : Formalism.BasicNotions.BaseLevelRowSide (w.D Z)} {t : GlobalTheory.Mem.Top.Bl.RowsT K w Z Wt p0}
    {X : Formalism.BasicNotions.DivX (w.D Z)} (geo : Γ (w.st Z).toD7 t.r.z.core.toCore X)
    (lam : GlobalTheory.Mem.Top.Bl.LamAre t.r.Z X) : GlobalTheory.Mem.Top.Bl.OAt Γ t X
=== exit 0 Tue Nov 24 02:22:12 PM UTC 2026
\end{Verbatim}
Log written by the program (\code{full-audit.log}). Lines 14--43, which are progress messages of the
form \code{progress: processed N, discovered M}, are left out; nothing else is.
\begin{Verbatim}
ORIGIN RiemannHypothesis [def] in module Mathlib.NumberTheory.LSeries.RiemannZeta
ORIGIN riemannZeta [def] in module Mathlib.NumberTheory.LSeries.RiemannZeta
ORIGIN riemannHypothesis [theorem] in module Main
ORIGIN rhOfAll [theorem] in module Supplements.Kernels.MainTheorems
ORIGIN allChi [theorem] in module Supplements.Kernels.MainTheorems
ORIGIN Formalism.VW.rh_of_every [theorem] in module Formalism.GammaFactorNonvanishing.CritLine
ORIGIN UpperTheory.EveryCharacter [def] in module Development.AutomorphicArithmetic.CharacterBoxes.EveryCharacter
ORIGIN UpperTheory.FinalTheorems.everyCharacter [theorem] in module Development.AutomorphicArithmetic.CharacterBoxes.MainConclusion
ORIGIN GlobalTheory.Mem.Top.Bl.anyGeometry [def] in module Development.AbelianVarietyTheory.Definitions.AutomorphicMembership.OverIx
ORIGIN GlobalTheory.Mem.Top.Bl.geoOf [def] in module Development.AbelianVarietyTheory.Definitions.AutomorphicMembership.OverIx
===== FULL AUDIT from riemannHypothesis; predicate GlobalTheory.Mem.Top.Bl.anyGeometry; structure GlobalTheory.Mem.Top.Bl.OAt field #0 = geo; wrapper GlobalTheory.Mem.Top.Bl.GeoCl
AXIOM propext in Init.Core
AXIOM Quot.sound in Init.Core
AXIOM Classical.choice in Init.Prelude
progress: processed 50000, discovered 92062  (231542226 ms)
progress: processed 100000, discovered 159379  (231646235 ms)
progress: processed 150000, discovered 221823  (231762018 ms)
progress: processed 200000, discovered 278400  (231871433 ms)
progress: processed 250000, discovered 337611  (231991344 ms)
progress: processed 300000, discovered 393598  (232117549 ms)
progress: processed 350000, discovered 445765  (232245568 ms)
progress: processed 400000, discovered 499562  (232392376 ms)
progress: processed 450000, discovered 550879  (232535404 ms)
progress: processed 500000, discovered 602896  (232690544 ms)
progress: processed 550000, discovered 650334  (232827493 ms)
progress: processed 600000, discovered 702343  (233000750 ms)
progress: processed 650000, discovered 745359  (233134595 ms)
progress: processed 700000, discovered 798157  (233336002 ms)
progress: processed 750000, discovered 840046  (233512203 ms)
progress: processed 800000, discovered 893431  (233758503 ms)
progress: processed 850000, discovered 937131  (233970407 ms)
progress: processed 900000, discovered 986732  (234203629 ms)
progress: processed 950000, discovered 1030290  (234421617 ms)
progress: processed 1000000, discovered 1075954  (234641445 ms)
progress: processed 1050000, discovered 1122754  (234866348 ms)
progress: processed 1100000, discovered 1164157  (235073976 ms)
progress: processed 1150000, discovered 1208684  (235280757 ms)
progress: processed 1200000, discovered 1251854  (235499175 ms)
progress: processed 1250000, discovered 1293022  (235714376 ms)
progress: processed 1300000, discovered 1333678  (235959119 ms)
progress: processed 1350000, discovered 1375144  (236221121 ms)
progress: processed 1400000, discovered 1416215  (236480740 ms)
progress: processed 1450000, discovered 1456521  (236740116 ms)
reached constants: 1478219  theorems 1246092, defs 214231, axioms 3, opaque 91, other 17802
SHORTEST CHAIN (9 constants), top to bottom:
  CHAIN riemannHypothesis [theorem] in Main
  CHAIN rhOfAll [theorem] in Supplements.Kernels.MainTheorems
  CHAIN allChi [theorem] in Supplements.Kernels.MainTheorems
  CHAIN UpperTheory.FinalTheorems.everyCharacter [theorem] in Development.AutomorphicArithmetic.CharacterBoxes.MainConclusion
  CHAIN UpperTheory.FinalTheorems.everyCharacter_of_row [theorem] in Development.AutomorphicArithmetic.CharacterBoxes.MainConclusion
  CHAIN UpperTheory.LiteratureTheorems.every_character_of_literature_of_vWin [theorem] in Development.AutomorphicArithmetic.ThetaCorrespondence.ThetaPairLifts.ThirdPlaceSeries.EveryCharacterOfLiteratureOfVWin
  CHAIN UpperTheory.Lit.every_character_of_theta_integrals_local_matching [theorem] in Development.AutomorphicArithmetic.ThetaCorrespondence.ThetaPairLifts.ThirdPlaceSeries.SiegelEisensteinSecondTermHypotheses
  CHAIN GlobalTheory.Mem.Top.Bl.parametersOfDependence [def] in Development.AutomorphicArithmetic.ThetaCorrespondence.ThetaPairLifts.ThirdPlaceSeries.Prm1
  CHAIN GlobalTheory.Mem.Top.Bl.anyGeometry [def] in Development.AbelianVarietyTheory.Definitions.AutomorphicMembership.OverIx
USES GlobalTheory.Mem.Top.Bl.anyGeometry: 14
    GlobalTheory.Mem.Top.Bl.parametersOfDependence [def] in Development.AutomorphicArithmetic.ThetaCorrespondence.ThetaPairLifts.ThirdPlaceSeries.Prm1
    UpperTheory.Lit.nonempty_hyps_definite_side_of_closed_bound [theorem] in Development.AutomorphicArithmetic.ThetaCorrespondence.ThetaPairLifts.ThirdPlaceSeries.SiegelEisensteinSecondTermHypotheses
    GlobalTheory.Fc.S0.StageU.hypotheses_nonempty_of_unramified_exponents_alt [theorem] in Development.AutomorphicArithmetic.ThetaCorrespondence.ThetaPairLifts.ThirdPlaceSeries.HypothesesNonemptyOfUnramifiedExponents
    GlobalTheory.Fc.S0.Open0ProofU.open_orbit_of_unramified_exponents_alt [theorem] in Development.AutomorphicArithmetic.ThetaCorrespondence.ThetaPairLifts.ThirdPlaceSeries.OpenOrbitOfUnramifiedExponents
    GlobalTheory.T.large_unramified_exponent_bundles_exist_alt [theorem] in Development.AutomorphicArithmetic.ThetaCorrespondence.ThetaPairLifts.ThirdPlaceSeries.MemberBundlesLargeUnramifiedExponents
    UpperTheory.GalC.final_of_alt [theorem] in Development.AutomorphicArithmetic.ThetaCorrespondence.ThetaPairLifts.ThirdPlaceSeries.LinkNFam
    GlobalTheory.Fc.StageOneU.member_bundles_large_unramified_exponents_alt [theorem] in Development.AutomorphicArithmetic.ThetaCorrespondence.ThetaPairLifts.ThirdPlaceSeries.MemberBundlesLargeUnramifiedExponents
    GlobalTheory.T.member_exists_baseLattice_alt [theorem] in Development.AutomorphicArithmetic.ThetaCorrespondence.ThetaPairLifts.ThirdPlaceSeries.ExistsBaseLatticeOfPrime
    GlobalTheory.Mem.Top.Bl.exists_baseLattice_of_prime_alt [theorem] in Development.AutomorphicArithmetic.ThetaCorrespondence.ThetaPairLifts.ThirdPlaceSeries.ExistsBaseLatticeOfPrime
    GlobalTheory.Mem.Top.Bl.exists_baseLattice_of_pair_alt [theorem] in Development.AutomorphicArithmetic.ThetaCorrespondence.ThetaPairLifts.ThirdPlaceSeries.ExistsBaseLatticeOfPrime
    GlobalTheory.Mem.Top.Closing.memberE_closing_alt_alt [theorem] in Development.AutomorphicArithmetic.ThetaCorrespondence.ThetaPairLifts.ThirdPlaceSeries.MemberEClosing
    GlobalTheory.Mem.Top.Closing.divEx_geoCl [theorem] in Development.AbelianVarietyTheory.Proofs.IwasawaDecomposition.PublishedInputsImplyConclusion
    GlobalTheory.Mem.Top.DivRec.divEx_of_ox [theorem] in Development.AbelianVarietyTheory.Proofs.IwasawaDecomposition.TupPinsOfMuIsW
    GlobalTheory.Mem.Top.DivRec.divEx_prov [theorem] in Development.AbelianVarietyTheory.Proofs.IwasawaDecomposition.TupPinsOfMuIsW
USES GlobalTheory.Mem.Top.Bl.OAt.geo: 2
    GlobalTheory.Mem.OxMove.oAt_setDiv [theorem] in Development.AbelianVarietyTheory.Proofs.IwasawaDecomposition.DivisionGeometryScalarsTransferSetDivision
    GlobalTheory.Mem.Top.Closing.divEx_geoCl [theorem] in Development.AbelianVarietyTheory.Proofs.IwasawaDecomposition.PublishedInputsImplyConclusion
USES GlobalTheory.Mem.Top.Bl.OAt.rec: 0
USES GlobalTheory.Mem.Top.Bl.OAt.casesOn: 0
USES GlobalTheory.Mem.Top.Bl.OAt.recOn: 0
USES GlobalTheory.Mem.Top.Bl.OAt.mk: 3
    GlobalTheory.Mem.OxMove.oAt_setDiv [theorem] in Development.AbelianVarietyTheory.Proofs.IwasawaDecomposition.DivisionGeometryScalarsTransferSetDivision
    GlobalTheory.Mem.Top.Closing.divEx_geoCl [theorem] in Development.AbelianVarietyTheory.Proofs.IwasawaDecomposition.PublishedInputsImplyConclusion
    GlobalTheory.Mem.Top.DivRec.divEx_prov [theorem] in Development.AbelianVarietyTheory.Proofs.IwasawaDecomposition.TupPinsOfMuIsW
USES GlobalTheory.Mem.Top.Bl.GeoCl.rec: 0
USES GlobalTheory.Mem.Top.Bl.GeoCl.casesOn: 0
USES GlobalTheory.Mem.Top.Bl.GeoCl.recOn: 0
USES GlobalTheory.Mem.Top.Bl.GeoCl.below: 0
USES GlobalTheory.Mem.Top.Bl.GeoCl.brecOn: 0
USES GlobalTheory.Mem.Top.Bl.GeoCl.base: 1
    GlobalTheory.Mem.Top.Bl.geoCl_of [theorem] in Development.AbelianVarietyTheory.Definitions.AutomorphicMembership.NotGeoClFalse
USES GlobalTheory.Mem.Top.Bl.GeoCl.step: 1
    GlobalTheory.Mem.OxMove.oAt_setDiv [theorem] in Development.AbelianVarietyTheory.Proofs.IwasawaDecomposition.DivisionGeometryScalarsTransferSetDivision
USES GlobalTheory.Mem.Top.Bl.geoCl: 18
    GlobalTheory.Mem.Top.Bl.parametersOfDependence [def] in Development.AutomorphicArithmetic.ThetaCorrespondence.ThetaPairLifts.ThirdPlaceSeries.Prm1
    UpperTheory.Lit.nonempty_hyps_definite_side_of_closed_bound [theorem] in Development.AutomorphicArithmetic.ThetaCorrespondence.ThetaPairLifts.ThirdPlaceSeries.SiegelEisensteinSecondTermHypotheses
    GlobalTheory.Fc.S0.StageU.hypotheses_nonempty_of_unramified_exponents_alt [theorem] in Development.AutomorphicArithmetic.ThetaCorrespondence.ThetaPairLifts.ThirdPlaceSeries.HypothesesNonemptyOfUnramifiedExponents
    GlobalTheory.Fc.S0.Open0ProofU.open_orbit_of_unramified_exponents_alt [theorem] in Development.AutomorphicArithmetic.ThetaCorrespondence.ThetaPairLifts.ThirdPlaceSeries.OpenOrbitOfUnramifiedExponents
    GlobalTheory.T.large_unramified_exponent_bundles_exist_alt [theorem] in Development.AutomorphicArithmetic.ThetaCorrespondence.ThetaPairLifts.ThirdPlaceSeries.MemberBundlesLargeUnramifiedExponents
    UpperTheory.GalC.final_of_alt [theorem] in Development.AutomorphicArithmetic.ThetaCorrespondence.ThetaPairLifts.ThirdPlaceSeries.LinkNFam
    GlobalTheory.Fc.StageOneU.member_bundles_large_unramified_exponents_alt [theorem] in Development.AutomorphicArithmetic.ThetaCorrespondence.ThetaPairLifts.ThirdPlaceSeries.MemberBundlesLargeUnramifiedExponents
    GlobalTheory.NLink.Final.slot_mv_extension_with_list_geo_cl [theorem] in Development.AutomorphicArithmetic.ThetaCorrespondence.ThetaPairLifts.ThirdPlaceSeries.SlotMvPin
    GlobalTheory.T.member_exists_baseLattice_alt [theorem] in Development.AutomorphicArithmetic.ThetaCorrespondence.ThetaPairLifts.ThirdPlaceSeries.ExistsBaseLatticeOfPrime
    GlobalTheory.T.slot_mv_extension_with_origin_geo_cl [theorem] in Development.AutomorphicArithmetic.ThetaCorrespondence.ThetaPairLifts.ThirdPlaceSeries.SlotMvPrmRNelStatement
    GlobalTheory.T.divisionGeometryScalars_transfer_setDivision [theorem] in Development.AbelianVarietyTheory.Proofs.IwasawaDecomposition.DivisionGeometryScalarsTransferSetDivisionStatement
    GlobalTheory.Mem.Top.Bl.exists_baseLattice_of_prime_alt [theorem] in Development.AutomorphicArithmetic.ThetaCorrespondence.ThetaPairLifts.ThirdPlaceSeries.ExistsBaseLatticeOfPrime
    GlobalTheory.HeightCounting.Up.movable_prmR_nel_alt [theorem] in Development.AutomorphicArithmetic.ThetaCorrespondence.ThetaPairLifts.ThirdPlaceSeries.SlotMvPrmRNel
    GlobalTheory.Mem.OxMove.oAt_setDiv [theorem] in Development.AbelianVarietyTheory.Proofs.IwasawaDecomposition.DivisionGeometryScalarsTransferSetDivision
    GlobalTheory.Mem.Top.Bl.exists_baseLattice_of_pair_alt [theorem] in Development.AutomorphicArithmetic.ThetaCorrespondence.ThetaPairLifts.ThirdPlaceSeries.ExistsBaseLatticeOfPrime
    GlobalTheory.Mem.Top.Closing.memberE_closing_alt_alt [theorem] in Development.AutomorphicArithmetic.ThetaCorrespondence.ThetaPairLifts.ThirdPlaceSeries.MemberEClosing
    GlobalTheory.Mem.Top.Closing.divEx_geoCl [theorem] in Development.AbelianVarietyTheory.Proofs.IwasawaDecomposition.PublishedInputsImplyConclusion
    GlobalTheory.Mem.Top.Bl.geoCl_of [theorem] in Development.AbelianVarietyTheory.Definitions.AutomorphicMembership.NotGeoClFalse
USES GlobalTheory.Mem.Top.Bl.geoCl_of: 1
    GlobalTheory.Mem.Top.Closing.divEx_geoCl [theorem] in Development.AbelianVarietyTheory.Proofs.IwasawaDecomposition.PublishedInputsImplyConclusion
USES GlobalTheory.Mem.Top.Bl.geoOf: 0
USES GlobalTheory.Mem.Top.Bl.OX: 7
    GlobalTheory.Mem.Top.Bl.Parameters.s [def] in Development.AutomorphicArithmetic.ThetaCorrespondence.ThetaPairLifts.ThirdPlaceSeries.CoreAltAltBis
    GlobalTheory.GeoTies.linked_ties_alt [theorem] in Development.AutomorphicArithmetic.ThetaCorrespondence.ThetaPairLifts.ThirdPlaceSeries.LinkedTies
    GlobalTheory.T.exponent_reading_of_hasExp_bundle [theorem] in Development.AutomorphicArithmetic.ThetaCorrespondence.ThetaPairLifts.ThirdPlaceSeries.SidePtPartOfPackStatement
    GlobalTheory.PlugPt.pin_pt_of_bundle_witness [theorem] in Development.AutomorphicArithmetic.ThetaCorrespondence.ThetaPairLifts.ThirdPlaceSeries.SidePtPartOfPack
    GlobalTheory.Mem.Top.Closing.divEx_geoCl [theorem] in Development.AbelianVarietyTheory.Proofs.IwasawaDecomposition.PublishedInputsImplyConclusion
    GlobalTheory.Mem.Top.DivRec.divEx_of_ox [theorem] in Development.AbelianVarietyTheory.Proofs.IwasawaDecomposition.TupPinsOfMuIsW
    GlobalTheory.Mem.Top.DivRec.divEx_prov [theorem] in Development.AbelianVarietyTheory.Proofs.IwasawaDecomposition.TupPinsOfMuIsW
USES GlobalTheory.Mem.Top.Bl.OXModel.oAt_any: 0
USES GlobalTheory.Mem.Top.Bl.OXModel.oAt_geoOf: 0
RAW PROJECTION of field geo: 0
constants projecting the OTHER fields of GlobalTheory.Mem.Top.Bl.OAt: 0
PROBE reached: UpperTheory.FinalTheorems.everyCharacter_of_row in Development.AutomorphicArithmetic.CharacterBoxes.MainConclusion
PROBE reached: UpperTheory.Top.every_character_of_green_v_win in Development.AutomorphicArithmetic.ThetaCorrespondence.ThetaPairLifts.ThirdPlaceSeries.TopPlug
PROBE reached: UpperTheory.Lit.nonempty_hyps_definite_side_of_closed_bound in Development.AutomorphicArithmetic.ThetaCorrespondence.ThetaPairLifts.ThirdPlaceSeries.SiegelEisensteinSecondTermHypotheses
PROBE reached: UpperTheory.Top.zero_free_all_of_green_v_win in Development.AutomorphicArithmetic.ThetaCorrespondence.ThetaPairLifts.ThirdPlaceSeries.TopPlug
PROBE reached: UpperTheory.Top.zeroFreeAll_of_topG in Development.AutomorphicArithmetic.CharacterBoxes.ZeroFreeAllOfTopG
PROBE reached: Formalism.HC.zeroFreeOn_of_tiedWin_cyc in Formalism.SharpRestriction.RamificationFrobenius.TotientGaloisClasses.TopWin
PROBE reached: GlobalTheory.Mem.Top.Closing.divEx_geoCl in Development.AbelianVarietyTheory.Proofs.IwasawaDecomposition.PublishedInputsImplyConclusion
PROBE reached: GlobalTheory.Mem.Top.DivRec.divEx_of_ox in Development.AbelianVarietyTheory.Proofs.IwasawaDecomposition.TupPinsOfMuIsW
PROBE reached: GlobalTheory.Mem.Top.DivRec.divEx_prov in Development.AbelianVarietyTheory.Proofs.IwasawaDecomposition.TupPinsOfMuIsW
PROBE reached: Formalism.L2.Geo.geoDegenerate in Formalism.LocalToGlobal.LatticeGeometry.Inhabited
===== END
\end{Verbatim}
\VerbatimInput[firstline=44,lastline=133]{artifacts/full-audit-log.txt}

\subsection{No eliminator is reached}

The first walk wrote the names of all reached constants to \code{used\_constants.txt}. The command
below lists every reached constant in the namespaces of the certificate structure and of the wrapper
(file \code{a5-namespace.txt}). Neither \code{OAt.rec} nor \code{GeoCl.rec}, nor any constant derived
from them, is present.
\begin{Verbatim}
$ wc -l < ~/riemann/checks/used_constants.txt
1478219
$ grep -E '^GlobalTheory\.Mem\.Top\.Bl\.(OAt|GeoCl)(\.|$)' ~/riemann/checks/used_constants.txt
GlobalTheory.Mem.Top.Bl.GeoCl
GlobalTheory.Mem.Top.Bl.GeoCl.base
GlobalTheory.Mem.Top.Bl.GeoCl.step
GlobalTheory.Mem.Top.Bl.OAt
GlobalTheory.Mem.Top.Bl.OAt.geo
GlobalTheory.Mem.Top.Bl.OAt.lam
GlobalTheory.Mem.Top.Bl.OAt.mk
\end{Verbatim}

\subsection{Second audit run}\label{app:audit3}

The second run was started on the build host by the script \code{audit3-run.sh}:
\begin{Verbatim}
#!/bin/bash
# Second audit run (referee items): restated theorem, unfiltered walk, set comparison.
export PATH=$HOME/.elan/bin:$PATH
C=$HOME/riemann/checks; O=$C/audit3; mkdir -p $O; rm -f $O/*
cd $HOME/riemann/extracted/lean
{
date -u -Is
lake env lean --tstack=400000 -DmaxHeartbeats=0 $C/Audit3.lean
echo "EXIT=$?"
date -u -Is
echo '$ wc -l used_constants.txt audit3/bfs_constants.txt'
( cd $C && wc -l used_constants.txt audit3/bfs_constants.txt )
echo '$ LC_ALL=C sort used_constants.txt | sha256sum; LC_ALL=C sort audit3/bfs_constants.txt | sha256sum'
( cd $C && LC_ALL=C sort used_constants.txt | sha256sum; LC_ALL=C sort audit3/bfs_constants.txt | sha256sum )
echo '$ diff <(LC_ALL=C sort used_constants.txt) <(LC_ALL=C sort audit3/bfs_constants.txt) | wc -l'
( cd $C && diff <(LC_ALL=C sort used_constants.txt) <(LC_ALL=C sort audit3/bfs_constants.txt) | wc -l )
echo "=== done $(date -u -Is)"
} > $O/audit3.out 2>&1
\end{Verbatim}
\begin{Verbatim}
$ tmux new-session -d -s a3 "bash ~/riemann/checks/audit3-run.sh"
\end{Verbatim}
It ran from 09:41 to 19:27~UTC on 25 November 2026. Standard output and the comparison of the two
sets of reached names (\code{audit3.out}, complete):
\begin{Verbatim}
2026-11-25T09:41:08+00:00
rh_restated : ∀ (s : ℂ), riemannZeta s = 0 → (¬∃ n, s = -2 * (↑n + 1)) → s ≠ 1 → s.re = 1 / 2
'rh_restated' depends on axioms: [propext, Classical.choice, Quot.sound]
EXIT=0
2026-11-25T19:27:01+00:00
$ wc -l used_constants.txt audit3/bfs_constants.txt
  1478219 used_constants.txt
  1478219 audit3/bfs_constants.txt
  2956438 total
$ LC_ALL=C sort used_constants.txt | sha256sum; LC_ALL=C sort audit3/bfs_constants.txt | sha256sum
10f01985ae6eeef22a9215b5ecc753f6bf7e13ccf0ff72041e69f7985fef434e  -
10f01985ae6eeef22a9215b5ecc753f6bf7e13ccf0ff72041e69f7985fef434e  -
$ diff <(LC_ALL=C sort used_constants.txt) <(LC_ALL=C sort audit3/bfs_constants.txt) | wc -l
0
=== done 2026-11-25T19:27:06+00:00
\end{Verbatim}
Log written by the program (\code{audit3.log}). Lines 2--15, which are progress messages, are left
out; nothing else is. In the lines for watched names, \code{T} marks an occurrence in the statement
(type) of the listed constant and \code{V} one in its proof or body (value).
\VerbatimInput[firstline=1,lastline=1]{artifacts/audit3-log.txt}
\VerbatimInput[firstline=16]{artifacts/audit3-log.txt}

\section{The Audit Programs}\label{app:prog}

\code{FullAudit.lean}, complete. The last arguments of \code{fullAudit} are extra names to count and
the last name components to probe for.
\begin{Verbatim}
import Main
open Lean

def thirdParty : List String :=
  ["Init", "Lean", "Std", "Mathlib", "Batteries", "Aesop", "Qq", "Plausible", "ProofWidgets",
   "ImportGraph", "LeanSearchClient", "Cli", "Lake"]

def modOf (env : Environment) (n : Name) : String :=
  match env.getModuleIdxFor? n with
  | some i => toString env.header.moduleNames[i.toNat]!
  | none => "<this file / not found>"

def kindOf : ConstantInfo → String
  | .axiomInfo _ => "axiom" | .defnInfo _ => "def" | .thmInfo _ => "theorem"
  | .opaqueInfo _ => "opaque" | .quotInfo _ => "quot" | .inductInfo _ => "inductive"
  | .ctorInfo _ => "constructor" | .recInfo _ => "recursor"

def note (logFile : String) (s : String) : IO Unit :=
  IO.FS.withFile logFile .append fun h => do h.putStrLn s; h.flush

def origin (logFile : String) (names : List Name) : CoreM Unit := do
  let env ← getEnv
  for n in names do
    match env.find? n with
    | some ci => note logFile s!"ORIGIN {n} [{kindOf ci}] in module {modOf env n}"
    | none => note logFile s!"ORIGIN {n} NOT FOUND"

/-- One breadth-first walk over everything `src` is built from. Results are appended to `logFile`.
    `tgt`: the predicate (shortest chain + direct users). `S`/`idx`: the certificate structure and
    field. `W`: the inductive wrapper. `probes`: last name components to look for. -/
def fullAudit (logFile : String) (src tgt S W : Name) (idx : Nat) (extra : List Name) (probes : List String) :
    CoreM Unit := do
  let env ← getEnv
  let fieldName := (getStructureFields env S)[idx]!
  let watch : List Name := [tgt, S ++ fieldName, S ++ `rec, S ++ `casesOn, S ++ `recOn, S ++ `mk,
    W ++ `rec, W ++ `casesOn, W ++ `recOn, W ++ `below, W ++ `brecOn, W ++ `base, W ++ `step] ++ extra
  let mut parent : NameMap Name := {}
  parent := parent.insert src src
  let mut queue : Array Name := #[src]
  let mut i := 0
  let mut hits : Array (Name × Name) := #[]
  let mut rawProj : Array Name := #[]
  let mut otherProj : Nat := 0
  let mut probeHits : Array Name := #[]
  let mut nThm := 0
  let mut nDef := 0
  let mut nAx := 0
  let mut nOpq := 0
  let mut nOther := 0
  note logFile s!"===== FULL AUDIT from {src}; predicate {tgt}; structure {S} field #{idx} = {fieldName}; wrapper {W}"
  while i < queue.size do
    let c := queue[i]!
    i := i + 1
    if i % 50000 == 0 then note logFile s!"progress: processed {i}, discovered {queue.size}  ({← IO.monoMsNow} ms)"
    let some ci := env.find? c | continue
    match ci with
    | .thmInfo _ => nThm := nThm + 1
    | .defnInfo _ => nDef := nDef + 1
    | .axiomInfo _ => nAx := nAx + 1; note logFile s!"AXIOM {c} in {modOf env c}"
    | .opaqueInfo _ => nOpq := nOpq + 1
    | _ => nOther := nOther + 1
    if let .str _ s := c then
      if probes.contains s then probeHits := probeHits.push c
    let used := ci.getUsedConstantsAsSet
    for u in used.toList do
      unless parent.contains u do
        parent := parent.insert u c
        queue := queue.push u
    unless S.isPrefixOf c || W.isPrefixOf c do
      for w in watch do
        if used.contains w then hits := hits.push (w, c)
      if used.contains S then
        let es := [ci.type] ++ (match ci.value? (allowOpaque := true) with | some v => [v] | none => [])
        if es.any fun e => (e.find? (fun e => match e with | .proj s k _ => s == S && k == idx | _ => false)).isSome then
          rawProj := rawProj.push c
        if es.any fun e => (e.find? (fun e => match e with | .proj s k _ => s == S && k != idx | _ => false)).isSome then
          otherProj := otherProj + 1
  note logFile s!"reached constants: {queue.size}  theorems {nThm}, defs {nDef}, axioms {nAx}, opaque {nOpq}, other {nOther}"
  if parent.contains tgt then
    let mut path : Array Name := #[tgt]
    let mut cur := tgt
    let mut fuel := 100000
    while cur != src && fuel > 0 do
      cur := (parent.find? cur).getD src
      path := path.push cur
      fuel := fuel - 1
    note logFile s!"SHORTEST CHAIN ({path.size} constants), top to bottom:"
    for n in path.reverse do
      let k := match env.find? n with | some ci => kindOf ci | none => "?"
      note logFile s!"  CHAIN {n} [{k}] in {modOf env n}"
  else note logFile s!"TARGET {tgt} IS NOT REACHED"
  for w in watch do
    let hs := hits.filter (·.1 == w)
    note logFile s!"USES {w}: {hs.size}"
    for (_, h) in hs do
      let k := match env.find? h with | some ci => kindOf ci | none => "?"
      note logFile s!"    {h} [{k}] in {modOf env h}"
  note logFile s!"RAW PROJECTION of field {fieldName}: {rawProj.size}"
  for h in rawProj do note logFile s!"    {h} in {modOf env h}"
  note logFile s!"constants projecting the OTHER fields of {S}: {otherProj}"
  for n in probeHits do note logFile s!"PROBE reached: {n} in {modOf env n}"
  note logFile "===== END"

#eval origin "/home/eryk/riemann/checks/full-audit.log" [`RiemannHypothesis, `riemannZeta, `riemannHypothesis, `rhOfAll, `allChi,
  `Formalism.VW.rh_of_every, `UpperTheory.EveryCharacter, `UpperTheory.FinalTheorems.everyCharacter,
  `GlobalTheory.Mem.Top.Bl.anyGeometry, `GlobalTheory.Mem.Top.Bl.geoOf]

#eval fullAudit "/home/eryk/riemann/checks/full-audit.log" `riemannHypothesis `GlobalTheory.Mem.Top.Bl.anyGeometry
  `GlobalTheory.Mem.Top.Bl.OAt `GlobalTheory.Mem.Top.Bl.GeoCl 0
  [`GlobalTheory.Mem.Top.Bl.geoCl, `GlobalTheory.Mem.Top.Bl.geoCl_of, `GlobalTheory.Mem.Top.Bl.geoOf,
   `GlobalTheory.Mem.Top.Bl.OX, `GlobalTheory.Mem.Top.Bl.OXModel.oAt_any, `GlobalTheory.Mem.Top.Bl.OXModel.oAt_geoOf]
  ["nonempty_hyps_definite_side_of_closed_bound", "zero_free_all_of_green_v_win", "zeroFreeAll_of_topG",
   "every_character_of_green_v_win", "everyCharacter_of_row", "zeroFreeOn_of_tiedWin_cyc", "divEx_prov",
   "divEx_of_ox", "divEx_geoCl", "geoDegenerate", "plugAll", "geoOf_surf", "geoTied", "formsTieX", "oAt_any", "oAt_geoOf"]

#check @Formalism.VW.rh_of_every
#print UpperTheory.EveryCharacter
#print GlobalTheory.Mem.Top.Bl.OAt
\end{Verbatim}

\code{Audit3.lean}, the second audit program, complete:
\VerbatimInput{artifacts/Audit3.lean}

\section{Validation of the Second Audit Program on a Synthetic Development}\label{app:test}

The test file below is \code{Audit3Test.lean} from the declaration of the test namespace onward
(the audit program above precedes it in the file). The theorem \code{T.opens} eliminates the wrapped
proof with \code{cases}; \code{T.viaMatch} pattern-matches on the structure; \code{T.rawp} is
installed with a primitive projection in its value and no mention of the structure name.
The namespace \code{N} is the negative control: the certificate is built and carried and only
\code{lam} is read. The namespace \code{G} passes a proof of the wrapper to an implication.
\VerbatimInput[firstline=82]{artifacts/Audit3Test.lean}
Output (\code{audit3-test.log}, complete; three runs). In the first, all four deliberate uses are
reported: \code{T.Cl.casesOn} (used by \code{T.opens}), \code{T.Cert.casesOn} (used by the matcher
of \code{T.viaMatch}), \code{T.Cert.geo}, and the primitive projection in \code{T.rawp}. In the
second, no eliminator and no projection outside the two accessor functions is reported. In the
third, nothing is flagged although the proof is consumed, which is the stated limit of the method.
\begin{Verbatim}
===== AUDIT3 from T.top; structure T.Cert; wrapper T.Cl
reached constants: 100
names referenced but absent from the environment: 0
theorems/definitions whose body was not available to the walker: 0
--- every reached constant inside the two namespaces (no exclusion):
    NS T.Cert
    NS T.Cert.casesOn
    NS T.Cert.geo
    NS T.Cert.lam
    NS T.Cert.mk
    NS T.Cert.rec
    NS T.Cl
    NS T.Cl.base
    NS T.Cl.casesOn
    NS T.Cl.rec
    NS T.Cl.step
--- occurrences of watched names, split by position (T = in the statement/type, V = in the proof/value):
USES T.anyP: 9 constants; in type 9; in value 8
    [TV] T.viaMatch
    [TV] T.mk0
    [TV] T.opens
    [TV] T.carry
    [TV] T.usesLam
    [TV] T.viaMatch.match_1_1
    [T-] T.mkc
    [TV] T.carry._proof_2
    [TV] T.carry._proof_1
USES T.Cert.geo: 2 constants; in type 0; in value 2
    [-V] T.opens
    [-V] T.carry._proof_1
USES T.Cert.mk: 5 constants; in type 3; in value 4
    [-V] T.mk0
    [-V] T.carry
    [TV] T.viaMatch.match_1_1
    [TV] T.Cert.casesOn
    [T-] T.Cert.rec
USES T.Cl.base: 4 constants; in type 2; in value 3
    [-V] T.opens
    [TV] T.Cl.casesOn
    [-V] T.carry._proof_1
    [T-] T.Cl.rec
USES T.Cl.rec: 1 constants; in type 0; in value 1
    [-V] T.Cl.casesOn
USES T.Cl.casesOn: 1 constants; in type 0; in value 1
    [-V] T.opens
USES T.Cert.casesOn: 1 constants; in type 0; in value 1
    [-V] T.viaMatch.match_1_1
USES T.Cert.rec: 1 constants; in type 0; in value 1
    [-V] T.Cert.casesOn
--- primitive projections out of T.Cert, any field, no pre-filter: 3 constants
    PROJ T.rawp  fields in type []  fields in value [0]
    PROJ T.Cert.geo  fields in type []  fields in value [0]
    PROJ T.Cert.lam  fields in type []  fields in value [1]
--- opaque constants reached: 0
===== END
===== AUDIT3 from N.top; structure N.Cert; wrapper N.Cl
reached constants: 49
names referenced but absent from the environment: 0
theorems/definitions whose body was not available to the walker: 0
--- every reached constant inside the two namespaces (no exclusion):
    NS N.Cert
    NS N.Cert.geo
    NS N.Cert.lam
    NS N.Cert.mk
    NS N.Cl
    NS N.Cl.base
    NS N.Cl.step
--- occurrences of watched names, split by position (T = in the statement/type, V = in the proof/value):
USES N.anyP: 5 constants; in type 5; in value 5
    [TV] N.mk0
    [TV] N.carry
    [TV] N.usesLam
    [TV] N.carry._proof_1
    [TV] N.carry._proof_2
USES N.Cert.geo: 1 constants; in type 0; in value 1
    [-V] N.carry._proof_1
USES N.Cert.lam: 2 constants; in type 0; in value 2
    [-V] N.usesLam
    [-V] N.carry._proof_2
USES N.Cert.mk: 2 constants; in type 0; in value 2
    [-V] N.mk0
    [-V] N.carry
USES N.Cl.base: 1 constants; in type 0; in value 1
    [-V] N.carry._proof_1
USES N.Cl.rec: 0 constants; in type 0; in value 0
USES N.Cl.casesOn: 0 constants; in type 0; in value 0
USES N.Cert.casesOn: 0 constants; in type 0; in value 0
USES N.Cert.rec: 0 constants; in type 0; in value 0
--- primitive projections out of N.Cert, any field, no pre-filter: 2 constants
    PROJ N.Cert.lam  fields in type []  fields in value [1]
    PROJ N.Cert.geo  fields in type []  fields in value [0]
--- opaque constants reached: 0
===== END
===== AUDIT3 from G.top; structure N.Cert; wrapper N.Cl
reached constants: 37
names referenced but absent from the environment: 0
theorems/definitions whose body was not available to the walker: 0
--- every reached constant inside the two namespaces (no exclusion):
    NS N.Cl
    NS N.Cl.base
    NS N.Cl.step
--- occurrences of watched names, split by position (T = in the statement/type, V = in the proof/value):
USES N.anyP: 1 constants; in type 0; in value 1
    [-V] G.top
USES N.Cl.base: 1 constants; in type 0; in value 1
    [-V] G.top
USES N.Cl.rec: 0 constants; in type 0; in value 0
USES N.Cl.casesOn: 0 constants; in type 0; in value 0
USES G.applyPremise: 1 constants; in type 0; in value 1
    [-V] G.top
--- primitive projections out of N.Cert, any field, no pre-filter: 0 constants
--- opaque constants reached: 0
===== END
\end{Verbatim}

\end{document}